\documentclass[numberedappendix]{emulateapj}
\usepackage{apjfonts}
\usepackage{lscape}
\usepackage{amsmath}
\pdfoutput=1
\shorttitle{}
\shortauthors{Mao et al.}

\begin{document}
\title{A Survey of Extragalactic Faraday Rotation at High Galactic Latitude: The Vertical Magnetic Field of the Milky Way towards the Galactic Poles}
\author{S. A. Mao,\altaffilmark{1,5} 
B. M. Gaensler,\altaffilmark{2}
M. Haverkorn,\altaffilmark{3}
E. G. Zweibel, \altaffilmark{4}
G. J. Madsen, \altaffilmark{2}
N. M. McClure-Griffiths,\altaffilmark{5}
A. Shukurov,\altaffilmark{6}
P. P. Kronberg\altaffilmark{7,8}
}

\altaffiltext{1}{Harvard-Smithsonian Center for Astrophysics, Cambridge, MA 02138; samao@cfa.harvard.edu}
\altaffiltext{2}{Sydney Institute for Astronomy, School of Physics, The University of Sydney, NSW 2006, Australia}
\altaffiltext{3}{ASTRON, Oude Hoogeveensedijk 4, 7991 PD Dwingeloo, The Netherlands}
\altaffiltext{4}{Department of Astronomy, University of Wisconsin, Madison, WI 53706}
\altaffiltext{5}{Australia Telescope National Facility, CSIRO, Epping, NSW 1710, Australia}
\altaffiltext{6}{School of Mathematics and Statistics, University of Newcastle, Newcastle upon Tyne, NE1 7RU, UK}
\altaffiltext{7}{Los Alamos National Laboratory, P.O. Box 1663, Los Alamos, NM 87545}
\altaffiltext{8}{Department of Physics, University of Toronto, 60 St. George Street, Toronto, M5S 1A7, Canada}

\begin{abstract}

We present a study of the vertical magnetic field of the Milky Way
towards the Galactic poles, determined from observations of Faraday
rotation toward more than 1000 polarized extragalactic radio sources
at Galactic latitudes $|b| \ge 77^\circ$, using the Westerbork Radio
Synthesis Telescope and the Australia Telescope Compact Array.  We
find median rotation measures (RMs) of $ 0.0 \pm 0.5$~rad~m$^{-2}$
and $+6.3\pm0.7$~rad~m$^{-2}$ toward the north and south Galactic
poles, respectively, demonstrating that there is no coherent vertical
magnetic field in the Milky Way at the Sun's position. If this is a global property of the Milky Way's magnetism, then the lack of symmetry across the
disk rules out pure dipole or quadrupole geometries for the Galactic
magnetic field.  The angular fluctuations in RM seen in our data
show no preferred scale within the range $\approx0\fdg1$ to
$\approx25^\circ$. The observed standard deviation in RM of
$\sim9$~rad~m$^{-2}$ then implies an upper limit of $\sim1$~$\mu$G on the strength of the random magnetic field in the warm ionized
medium at high Galactic latitudes.

\end{abstract}


\section{Introduction}
\label{section:introduction}

 Large scale coherent magnetic fields are observed in our Milky Way and in external galaxies \citep{beck2008b,beck2008a}; these fields play crucial roles in many astrophysical processes in the interstellar medium (ISM) --- they help to exert pressure to balance ordinary matter against gravity and trigger star formation, they are also responsible for the confinement of cosmic rays, and they can regulate and trace a large scale galactic wind. Therefore, to better understand galaxy evolution, it is necessary to investigate the structure, origin and evolution of galactic magnetic fields.  The primordial theory and the dynamo theory are the two possible explanations for the existence of a galactic scale magnetic field. These two theories make certain predictions on the symmetry of the large scale magnetic field with respect to the rotation axis and the mid-plane of the galaxy \citep[e.g.,][]{beck1996}. The characterization of the overall magnetic field geometry of a galaxy allows one to distinguish between various primordial and dynamo models. 

Studies of the large scale magnetic field geometry in the Milky Way have been mainly focused on the disk field symmetry with respect to the rotation axis \citep[e.g.,][]{han2006,brown2007} even though sight lines towards high Galactic latitude may be less turbulent and have less tangled fields than are seen in the Galactic plane and hence global field patterns can be identified more easily. The strength and the symmetry of the vertical, azimuthal and radial components of the Galactic magnetic field across the plane can provide us with unique insights on the mechanisms which maintain the Galactic magnetic field. For example, a reversal across the Galactic plane, in the azimuthal component of the large-scale magnetic field but not in its vertical component would indicate a field of dipolar structure, which could result from weak differential rotation (near solid body rotation) \citep{ferriere2005}; or a substantial primordial field \citep{zweibel1997}. On the other hand, if  there is a reversal in the vertical but not the azimuthal component across the mid-plane, it would indicate a field of quadrupolar structure resulting from dynamo action due to differential rotation in the Galactic disk \citep{zweibel1997}. The dynamo theory predicts a weak vertical field compared to the horizontal field on galactic scales. If the measured large-scale vertical field strength is substantially larger, it could imply a  primordial component to the Galactic magnetic field \citep{ruzmaikin1988}. The ratio of the vertical to horizontal field strength also regulates the confinement of cosmic rays to the Galactic disk, which can help us better understand the disk-halo interaction in the Milky Way.

Some information on the large scale structure of the Galactic magnetic field can come from optical starlight polarization and radio synchrotron polarization \citep[see][for a summary]{beck1996,beck2008b,beck2008a}. A series of papers: \cite{berdyugin2000,berdyugin2001a,berdyugin2001b,berdyugin2004} found that towards the south Galactic pole, the polarization of stars traces a field orientation along $\ell$ of 80$^\circ$ which is parallel to the local spiral arm field, while the same field direction could not be traced towards the North Galactic pole. Most of our knowledge of the geometry of the large-scale Galactic magnetic field to date comes from the Faraday rotation measure of distant extragalactic radio sources (EGSs) and pulsars. Faraday rotation is a birefringence effect when linearly polarized light travels through a magnetized media. The plane of the polarization rotates through an angle $\Delta$$\psi$ (in radians) given by
\begin{equation}
\Delta\psi= {\rm RM}  \lambda^{2}, 
\end{equation}
where $\lambda$ is the wavelength of the radiation measured in meters and RM is the rotation measure, defined as  the integral of the line of sight magnetic field $B_{\parallel}$ (in $\mu$G) weighted by the thermal electron density $n_{e}(l)$  (in cm$^{-3}$) over a line-of-sight line element $d$$\it{l}$ (in pc).
\begin{equation}
{\rm RM} =0.812 \int ^{observer}_{source} {n_{e}(l){B_{\parallel} (l)}} dl~~~\rm{rad~m^{-2}}. 
\label{eq:rmdef}
\end{equation}
The sign of the RM gives the direction of the line of sight component of the average field: a negative RM represents a magnetic field whose line of sight component is directed away from us. With independent knowledge of the thermal electron density, one can determine the average line-of-sight magnetic field strength.  Faraday rotation is complementary to other measurement techniques  since RMs provide the direction of the magnetic field, while most other techniques (apart from Zeeman splitting) provide the field orientation, but not its direction.

High latitude magnetic fields in the Milky Way have been investigated using RMs of EGSs by \cite{morris1964}, \cite{andreasyan1980}, \cite{andreassian1988} and \cite{han1994}. They all found a strong antisymmetric RM pattern between the northern and southern Galactic hemispheres. These authors attribute the pattern to a horizontal magnetic field that reverses direction above and below the Galactic plane. \cite{han1994} and \cite{han1999} estimated the local vertical magnetic field strength to be  $\sim$ 0.3 $\mu$G, pointing from the south to the north Galactic pole assuming a priori a dipolar field geometry. A  wavelet analysis on an all sky RM catalogue presented by \cite{frick2001} found that the RM distribution at the largest scales is shifted  to negative latitudes, indicative of  a stronger magnetic field in the southern hemisphere and hence the possible existence of an antisymmetric\footnotemark[1]\footnotetext[1]{In this paper, the terms symmetric and antisymmetric are used to describe the vertical and horizontal components of the Galactic magnetic field instead of the total magnetic field vector itself.} halo magnetic field\footnotemark[2]\footnotetext[2]{Throughout the paper, we refer to halo fields as non-disk fields that are in regions with sufficient diffuse interstellar electrons to produce Faraday rotation}. Unfortunately, local distortions such as that of a Parker instability loop cannot be ruled out.  A dipolar halo field is also suggested by \cite{sun2008}, who demonstrated that an oppositely directed torus-like halo field above and below the Galactic plane and a symmetric disk field provide a reasonable fit to the latitude extension of the Canadian Galactic Plane Survey (CGPS) RM measurements, but it is unclear if this model can predict RMs that are consistent with observations for all ranges of $\ell$ and $b$. 

\cite{jansson2009} found that the halo field might not be anti-symmetric globally since a halo field that reverses across the plane only towards the inner Galaxy provides a better fit to the WMAP 22GHz all sky polarization map and an all-sky catalogue of  $\sim$ 1,400 RMs of EGSs. Recently, \cite{taylor2009} used EGS RMs calculated from the NRAO VLA Sky Survey (NVSS) catalogue, a 1.4GHz continuum survey of the entire sky north of DEC=$-$40$^\circ$,  to derive a vertical field strength of  0.30 $\pm$ 0.03 $\mu$G for $b$$<$0$^\circ$ pointing from the south pole towards the north pole and a field strength of 0.14 $\pm$ 0.02 $\mu$G for $b$$>$0$^\circ$ pointing from the north towards the south pole. This new result is incompatible with the simple dipolar field model concluded in earlier studies. 

The current knowledge of the parity of the Galactic magnetic field is based on the sparse all-sky RM measurements and their potential unreliable values because they were derived using polarization angles at only a few and sometimes widely separated wavelengths. Moreover, most conclusions were drawn using low-latitude RMs which are likely to be contaminated by  turbulence and tangled fields in the Galactic disk. In this paper, we present an accurate rotation measure survey of more than 1,000  polarized extragalactic sources previously identified in the NVSS catalogue towards both the north and the south Galactic poles at a sampling density of approximately 1 source/deg$^2$. Our goal is to measure the magnetic field structure at high latitude in great details and to study the symmetry of the vertical Galactic magnetic field across the disk plane. In \S~\ref{section:observations}, we describe data acquisition and reduction procedures. The RM extraction procedure is outlined in \S~\ref{section:rmcomputation}. The results are presented in \S~\ref{section:results}. We discuss different possible origins of the measured RM pattern in \S~\ref{section:discussion} and derive the corresponding magnetic field properties, including both the coherent (\S~\ref{section:largescalefield}) and the random magnetic field strength (\S~\ref{section:random}).

\section{Observations and Data Reduction}
\label{section:observations}

To measure the vertical magnetic field at the location of the Sun, RMs along many sight lines towards the Galactic poles are desired. Since there is only one pulsar with known RM towards each of the Galactic poles (see \S~\ref{subsection:vertical}), one has to rely on RMs of EGSs to measure the high latitude Galactic magnetic fields. To avoid the inefficiency of conducting a blind survey, we applied appropriate filters to the NVSS catalogue \citep{condon1998} to yield a target list consists of highly polarized compact EGSs near the Galactic poles, towards which one can measure RMs. In order to determine RMs as accurately and precisely as possible, we chose to observe point-like extragalactic sources with a signal to noise ratio in linear polarization of at least 8. To allow for a reasonable observing time per source, we set our polarized flux threshold at 4 mJy. A rough estimation shows that a vertical magnetic field of strength 0.2 $\mu$G  and a total column density of thermal electrons towards the Galactic poles of $\sim$ 25 pc cm$^{-6}$  \citep{gaensler2008} would produce a  RM of  magnitude 4 rad m$^{-2}$. Since the intrinsic RM dispersion of extragalactic sources is roughly 15 rad m$^{-2}$ \citep{gaensler2005}, one needs approximately 500 sources to detect the predicted RM signal at a decent signal-to-noise level. The polarized flux cutoff at 4 mJy and the need of $\sim$ 500 EGSs requires us to select sources down to a Galactic latitude $|$$b$$|$ of 77$^\circ$ in the NVSS catalogue. 

\subsection{North Galactic Cap}

RM of EGSs towards the North Galactic pole were acquired at the Westerbork Synthesis Radio Telescope (WSRT) over the period 2006 July 23rd to August 20th using the  9A=96m  array configuration spanning baselines from 96 m to 2760 m; with 8 20-MHz frequency bands (each band is composed of 64 channels of width 0.3125 MHz) centered on each of 1381MHz and 1713MHz. Additional observations were made on 2009 April 7th to 8th with the mini-short configuration, spanning baselines from 96 m to  2723 m centered on only 1381MHz. 

We selected 500 EGSs with polarized intensity greater than 4 mJy at $b$ $>$ $+$77$^\circ$ from the NVSS catalogue. For each source, we obtain 30 cuts of 10 seconds at both 1381MHz and 1713MHz (except for sources observed in the additional session in 2009), resulting in an integration time of 5 minutes per source for each band and hence a total observing time of rougly 90 hours. Upon inspection of the data, we found that the IF centered around 1713MHz was heavily contaminated by radio frequency interference (RFI) and hence we decided to only use the IF centered around 1381MHz. The overall sensitivity of each target is $\sim$  0.1 mJy-beam$^{-1}$.  At least one of the standard absolute flux calibrators 3C286, 3C147 or CTD93 was observed before and after each 12 hour observing run to allow for amplitude and phase calibration. Bandpass and leakage correction was determined by  observing an unpolarized source, either 3C147 or CTD93.  A source with known polarization, one or both of 3C286 and 3C138 was observed for absolute polarization angle calibration. 

The NRAO's Astronomical Image Processing System (AIPS) was used for data edition and calibration. AIPS is designed to process interferometric data for telescopes with circular feeds while the WSRT consists of orthogonal linear feeds \citep{weiler1973}. Therefore, to properly calibrate and reduce WSRT data in AIPS requires  various non-standard procedures, such as relabeling the Stokes parameters. These calibration steps are described in details in the AIPS reduction cookbook for WSRT data\footnotemark[3]\footnotetext[3]{http://www.astron.nl/radio-observatory/astronomers/analysis-wsrt-data/analysis-wsrt-dzb-data-classic-aips/analysis-wsrt-d}. After calibration, flagging and rebinning the 512 0.3125MHz-wide channels results in 16 10-MHz wide channels centered around 20 cm. The u-v data were then exported to FITS format so that imaging could be carried out in the MIRIAD package  \citep{sault2003}. For each pointing and for each of the 16 10-MHz wide frequency channel, maps of Stokes parameters $Q$ and $U$ were made using the INVERT task in MIRIAD with natural weighting to maximize sensitivity. These maps were then deconvolved using the CLEAN algorithm with a threshold of 5 $\sigma$. A restored map for each pointing, for each Stokes $Q$ and $U$ at each of the 16 channels around  20 cm were made. This results in a total of $\sim$ 16,000 channel maps, each with a sensitivity of roughly 0.4 mJy-beam$^{-1}$ and resolution of $\sim$ 0.5'. A linearly polarized intensity (PI) map, corrected for positive bias was made for each source. The brightest polarized pixel of each EGS is identified and its Stokes $Q$ and $U$ values across the frequency band were extracted for RM determination. 

\subsection{South Galactic Cap }

RMs of EGSs towards the south Galactic pole were acquired at the Australia Telescope Compact Array (ATCA) over the period 2006 June 23rd to July 4th, using the 1.5D array configuration spanning baselines from 107.1 m to 1439 m (without antenna 6), and 2008 June 22nd and June 25th, using the  EW352 array configuration spanning baselines from 30.6 m to 352 m (without antenna 6). All ATCA data consist of 32 adjacent frequency channels each of bandwidth 4 MHz centered on each of 1384 MHz and 2368 MHz.

The standard primary flux and bandpass calibrator PKS B1934$-$638, whose flux at 1384 (2368) MHz was assumed to be 14.94 (11.14) Jy, was observed at the beginning and the end of each observation. A secondary calibrator (PKS 2339$-$353, PKS 0023$-$26, PKS 0010$-$401 or PKS 0153$-$410) was observed every hour and was used to correct for polarization leakages and to calibrate the time-dependent antenna gains \citep{sault1991}. We have applied the same EGS selection criteria as that towards the north pole and selected  533 sources towards the south Galactic pole. Allowing $\sim$ 15 minutes per pointing results in a total observing time of approximately 150 hours. The overall sensitivity of each field is $\sim$ 0.3 mJy-beam$^{-1}$.  Using the same pipeline that processes the north Galactic pole data, we have imaged  EGSs towards the south Galactic pole, resulting in images of resolution $\sim$ 0.5', similar to that of the north cap data. A total of  $\sim$ 26,500 Stokes $Q$ and $U$ channel maps (13 channels around 13 cm and 12 channels around 20 cm) were produced, each with a sensitivity of roughly 1 mJy-beam$^{-1}$.

\subsection{H$\alpha$ Observations}
\label{section:halpha}
The warm ionized medium (WIM), consisting of diffuse ionized gas, is a major constituent of the interstellar medium in the Milky Way. The WIM can be probed using the H$\alpha$ line and other optical recombination lines. The Wisconsin H$\alpha$ Mapper (WHAM) is a 1$^\circ$-angular resolution, velocity resolved survey of the H$\alpha$ line in the northern (completed) and the southern (in progress) sky \citep{haffner1999}. H$\alpha$ measurements together with dispersion measure (DM)
\begin{equation}
{\rm DM} =  \int ^{L}_{0} {n_{e}}{dl} =  \overline n_{e} L~~\rm{pc~cm^{-3}}
\end{equation}
 of pulsars along a line of sight can provide information on the clumpiness of the WIM \citep[see for example,][]{reynolds1991,gaensler2008}. The observed H$\alpha$ intensity in Rayleighs (1~R = 10$^6$ photons per 4$\pi$ steradian = 2.42 $\times$ 10$^{-7}$ ergs cm$^{-2}$ s$^{-1}$  sr$^{-1}$) towards the Galactic caps can be converted into emission measure (EM) assuming a constant electron temperature around 10,000K and negligible extinction at high Galactic latitudes \citep[$\le$ 0.02 mag][]{schlegel1998} using the following equation \citep{reynolds1991}
\begin{equation}
{\rm EM} = 2.75 (T/10^4)^{0.9} I_{H\alpha}.
\end{equation}
The EM along a sight line is defined as
\begin{equation}
\label{eq:em}
{\rm EM} = \int ^{L}_{0} {n_{e}^2 }{dl} = \overline{n_{e}^2} L~~\rm{pc~cm^{-6}}, 
\end{equation}
where $\overline{n_e^2}$ is the average of the electron density squared along the total path length $L$. 

The WHAM survey has full coverage of the north Galactic cap ($b$$>$+77$^\circ$) and currently a $\sim$  50$\%$  coverage of the south Galactic cap ($b$$<-$77$^\circ$). The average surface brightness is about 0.5 R(EM $\sim$ 1 pc~cm$^{-6}$) towards the caps. We have verified that there are no discrete HII regions at $|$$b$$|$$\ge$77$^\circ$. The WHAM data towards the cap regions are used to search for correlations between EM and RM in \S~\ref{subsection:NPS}.

\section{RM computation}
\label{section:rmcomputation}

Traditionally, the RM of an EGS is calculated by a linear least square fit of the unwrapped polarization position angle $\psi$
\begin{equation}
\psi = \frac {1}{2} \tan^{-1}\frac{U}{Q}
\end{equation}
as a function of the square of the wavelength. However, since the polarization angle can wrap around n$\pi$, there exists multiple RMs that provide equally good least square fits to the observed angle versus $\lambda$$^2$ relation. 
We have used the Faraday rotation measure synthesis, first presented by \cite{burn1966} and more recently illustrated in \cite{brentjens2005} and \cite{heald2009}, to obtain reliable RMs of the observed EGSs. This method is equivalent to taking the Fourier transform of the complex polarization vector measured at each frequency channel. The RM Synthesis method does not operate with polarization position angles and hence does not lead to the n$\pi$ ambiguity problem. Also, non-Gaussian statistics of polarization angles can be avoided. Moreover, this method can resolve multiple RM components along the line of sight or within the telescope beam, for which case the angle versus $\lambda$$^2$ fit would not give a sensible result. Backends of the ATCA and the WSRT make the implementation of RM synthesis possible as Stokes Q and U information can be obtained  in numerous neighboring frequency channels. 

We have performed RM Synthesis on the observed EGSs following the method described in detail by \cite{brentjens2005}. Similar to an aperture synthesis experiment where large gaps in the uv-plane causes high-level sidelobes in the image plane, incomplete wavelength coverage of the RM experiment leads to sidelobes in the resulting Faraday depth spectra. 

This can cause serious problems when attempting to extract the correct RM. To enable accurate extraction of both the RM and the polarized flux, we have deconvolved the complex Faraday depth spectrum using the CLEAN algorithm as described by \cite{brentjens2007} \citep[also see the Appendix in][]{heald2009}. We have adopted a gain factor of 0.1 and we stop CLEANing once the residual peak falls below 4 times the noise level in the final spectrum. The value of RM and the polarized intensity of the EGS is calculated by fitting a parabola to the peak(s) of the deconvolved Faraday depth power spectrum. The precision of a RM measurement is determined by the signal-to-noise of the polarization detection and the FWHM of the rotation measure transfer function (Equation (61) in  \cite{brentjens2005}). The measurement uncertainty of an RM with a debiased peak polarized intensity\footnotemark[4]\footnotetext[4]{The polarized intensity is debiased to the first order by subtracting the noise in the Faraday depth spectrum from the measured peak polarized intensity in quadrature \citep{simmons1985}.}  in the Faraday depth spectrum of PI$_{\rm peak}$, a rms error in the CLEANed Faraday depth spectrum $\sigma$  and a dirty beam FWHM of $\delta \phi$ rad m$^{-2}$ is
\begin{equation}
\Delta {\rm RM} = \frac {\delta \phi} { 2  {\rm{PI}_{peak}} / \sigma}. 
\end{equation}
The FWHM of the dirty beam is 343 rad m$^{-2}$ for the WSRT observations while the value for the ATCA observations is 99 rad m$^{-2}$. The ATCA observation has relatively low sensitivity but small $\delta \phi$, while the WSRT observation has better sensitivity but larger $\delta \phi$. This results in comparable RM errors for both the northern and southern data sets. We have verified, using several sources with steep/ shallow spectra in our sample, that the spectral index has little effect on the location of the Faraday dispersion function peak \citep{brentjens2005}. We therefore choose not to include spectral index effect in our RM computation/ deconvolution procedure.

\begin{figure}
\centering
\includegraphics[width=0.4\textwidth]{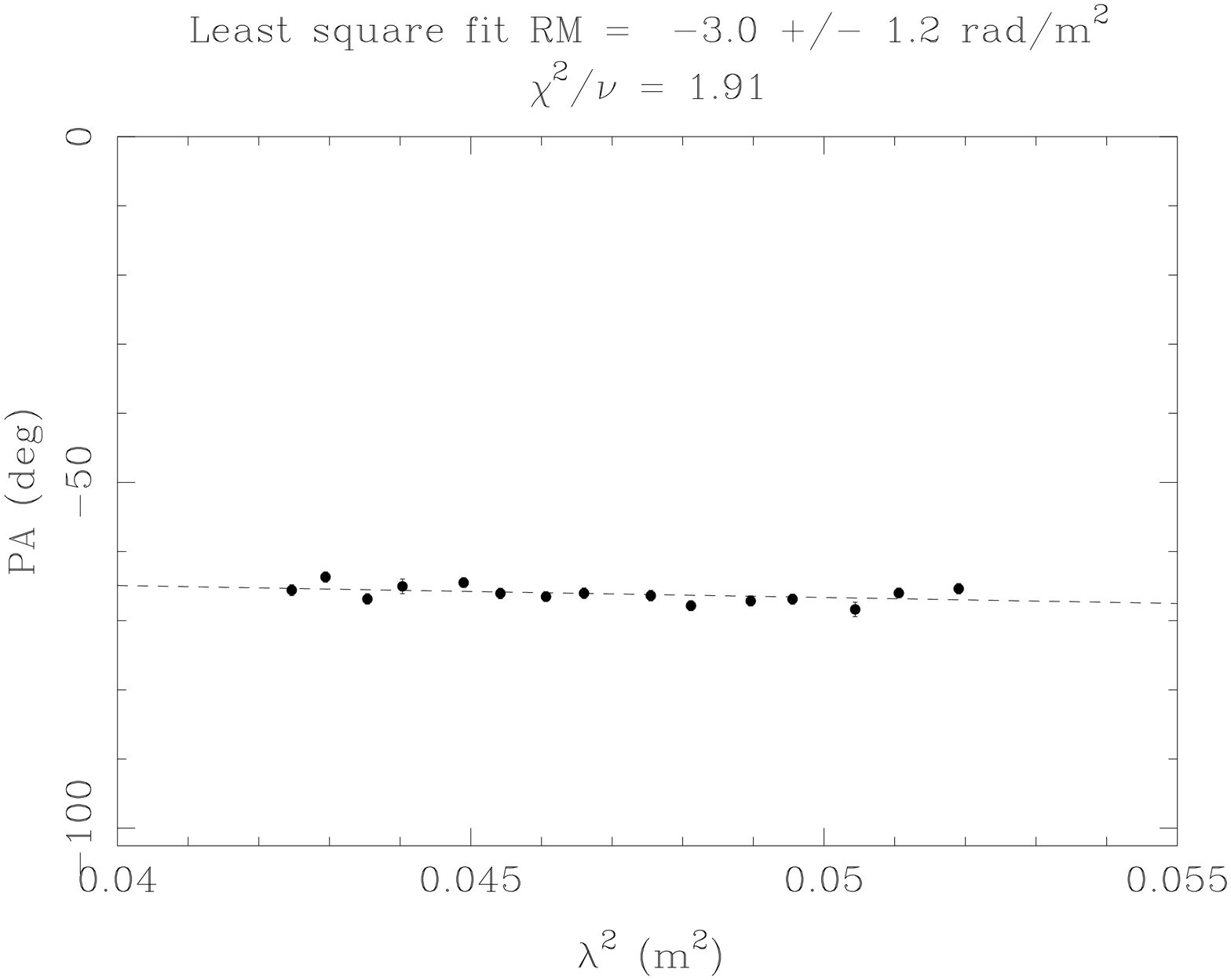}
\hspace{.05in}
\includegraphics[width=0.4\textwidth]{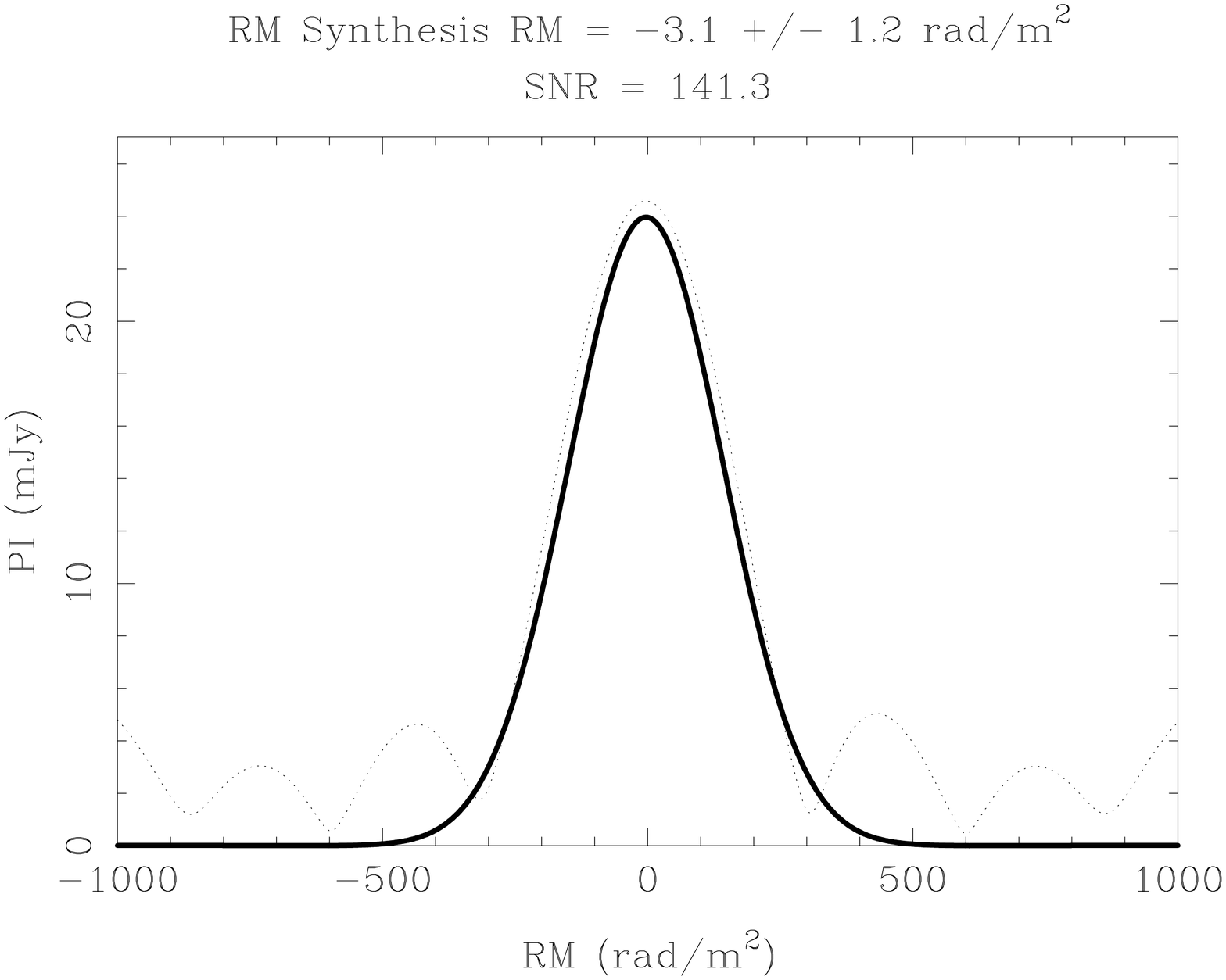}
\vspace{.05in}
\includegraphics[width=0.4\textwidth]{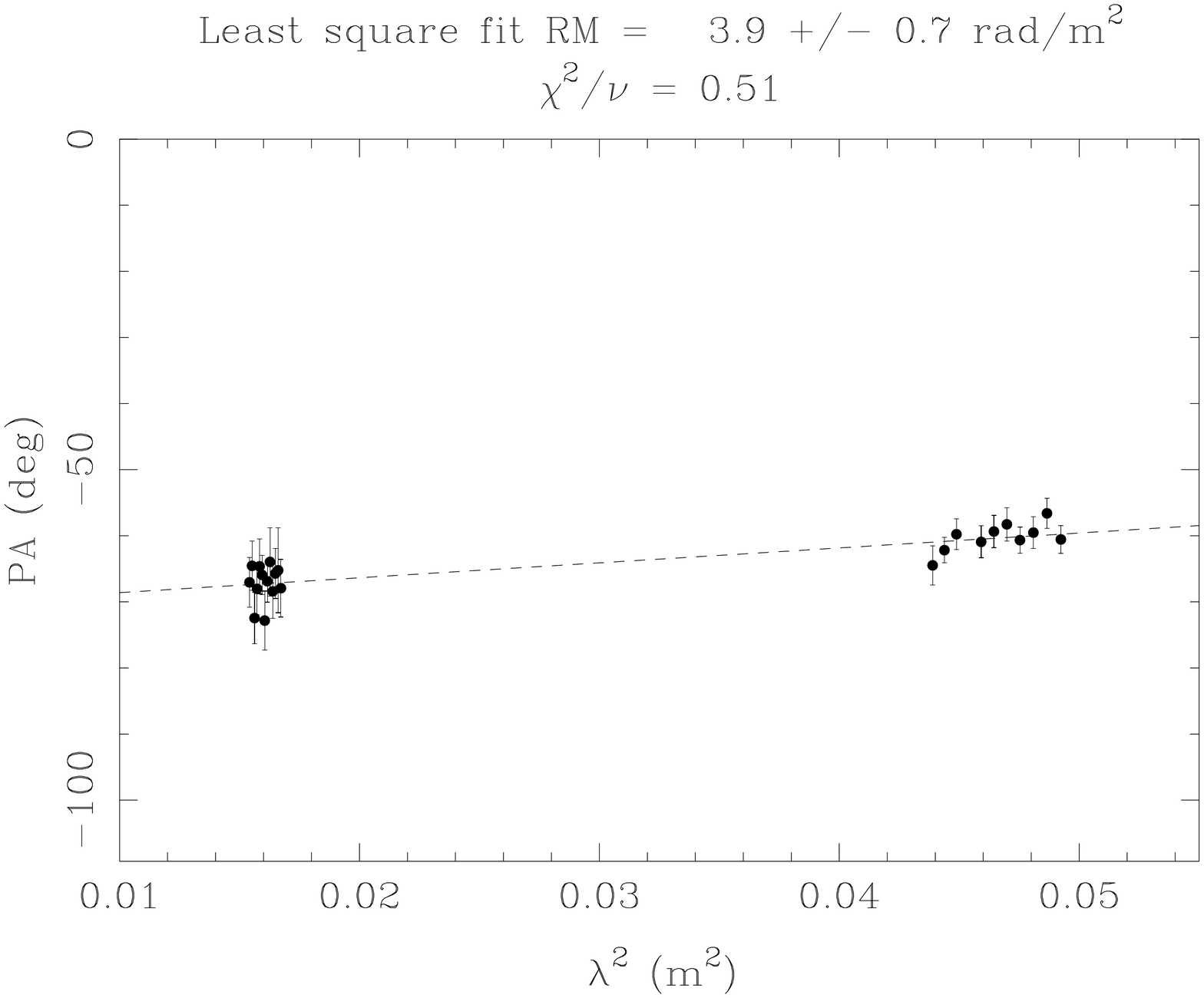}
\hspace{.05in}
\includegraphics[width=0.4\textwidth]{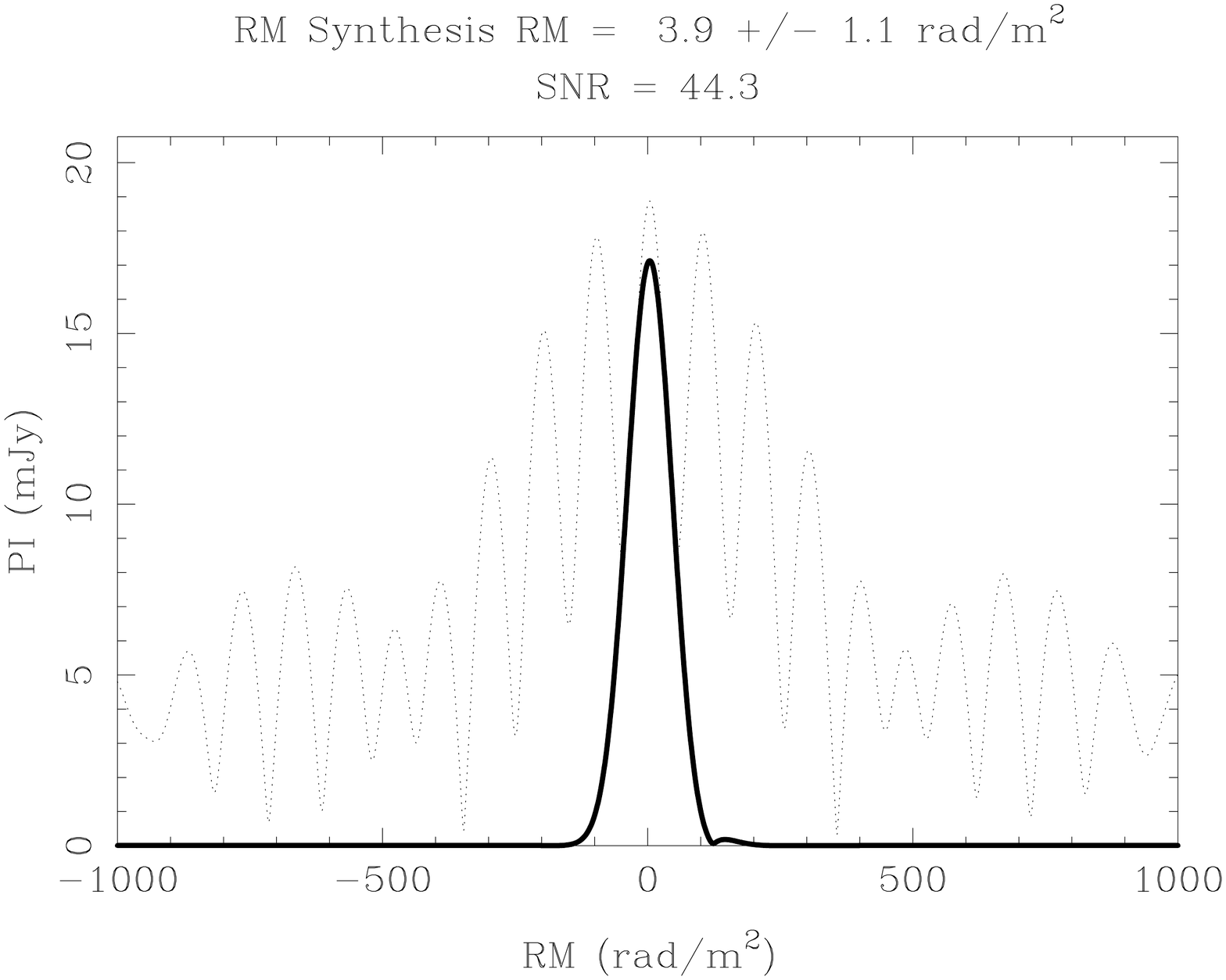}
\epsscale{1}

\caption{The least square fit and RM synthesis results for an EGS towards the north (RA=12:43:41.17, DEC=+31:24:17.88) in the top row and an EGS towards the south (RA=00:31:02.73, DEC=$-$22:07:06.50) in the bottom row. The left panel shows the position angle of linear polarization versus $\lambda^2$ relation for each source, with the best fit plotted as a dotted line. The right panel is the amplitude of the linear polarization as a function of Faraday depth, the thin dotted line is the dirty spectrum and the thick solid line is the deconvolved spectrum.}
\label{fig:rmfit}
\end{figure}

We have followed the above steps to derive RMs of EGSs towards the north Galactic cap using the 20 cm WSRT data. For sight lines towards the south Galactic cap, RMs were first computed using only the 20 cm ATCA data. If the source appears to have a well behaved Faraday depth spectrum with a single RM component, we then add in the 13 cm ATCA data to achieve better RM precision. This step requires the assumption that the same RM component dominates at both 13 and 20 cm. To minimize the potential systematics (such as calibration errors) introduced when we calculate the RMs using both bands, we have rejected sources that have RM values derived from the 20 cm only data that are at 2$\sigma$ level different from those obtained using both bands. We have also discarded sources with polarization detection signal-to-noise ratio below 8 and  with polarized fraction greater than 30 $\%$\footnotemark[5]\footnotetext[5]{Sources with unusually high polarized fraction might be of Galactic origin}. The least square fit and Faraday depth spectra plots of two example EGSs are shown in Fig~\ref{fig:rmfit}. Since the frequency setup for the ATCA observations leaves a larger gap in $\lambda^2$ space than the WSRT observations, it leads to higher side lobe levels seen in the ATCA Faraday depth spectrum. For bright polarized sources (as the two example EGSs), the least square fit and the RM synthesis method should yield converging results. 

As the Faraday rotation phenomenon is a line-of-sight effect, the observed RM is the sum of all the RM contributions along a sight line. This includes the intrinsic RM which originates in the magneto-ionic environment at the radio source; the RM produced in the intergalactic medium (IGM) or any intervener(s); that produced by the Milky Way's large scale and small scale field; as well as that produced by the ionosphere. The RM contribution from the IGM  is negligible -- cosmological Faraday rotation measure has an estimated upper limit of  $\sim$ 2 rad m$^{-2}$ assuming that the cosmological magnetic field and electron density is homogenous  \citep{vallee1990}. On the other hand, if these two quantities vary on scales much smaller than the Hubble scale, then no cosmological imprint to the all sky RM is expected as it would average out to be zero.  

As Global Positioning System (GPS) data were not available to monitor the real time ionospheric activity during the observations at the ATCA and the WSRT, we can only rely on theoretical models to estimate the contribution of ionospheric RM to the observed RMs. We note that theoretical predictions do not take into account of sudden and temporal ionospheric changes. We have used the International Geomagnetic Reference Field (IGRF) Model\footnotemark[6]\footnotetext[6]{Available at \url{http://omniweb.gsfc.nasa.gov/vitmo/igrf$\_$vitmo.html}} \citep{macmillan2005},
along with the International Reference Ionosphere (IRI) 2007\footnotemark[7]\footnotetext[7]{Available at \url{http://omniweb.gsfc.nasa.gov/vitmo/iri$\_$vitmo.html}} \citep{bilitza2008} to estimate the ionospheric RM during the observations. It has been found that the ionospheric RM above the ATCA is negative (geomagnetic field points away from the center of the earth) and that above the WSRT is positive (geomagnetic field points towards the center of the earth) with a magnitude $\le$ 1 rad m$^{-2}$, which is consistent with the estimation reported by \cite{tinbergen1996}. The variation of ionospheric RM along a certain RA and DEC over the course of the observation is also $\le$ 1 rad m$^{-2}$. This can be verified by computing the RM of a bright polarized EGS as a function of time over the observing run. We find that the RM stays constant as a function of time within its error. To account for the RM uncertainties introduced by  Faraday rotation from the ionosphere, we have added a systematic error of 1 rad m$^{-2}$ in quadrature to the RM measurement error for all sources.

\section {Results}
\label{section:results}

\begin{figure}
\centerline{
\includegraphics[width=0.8\textwidth]{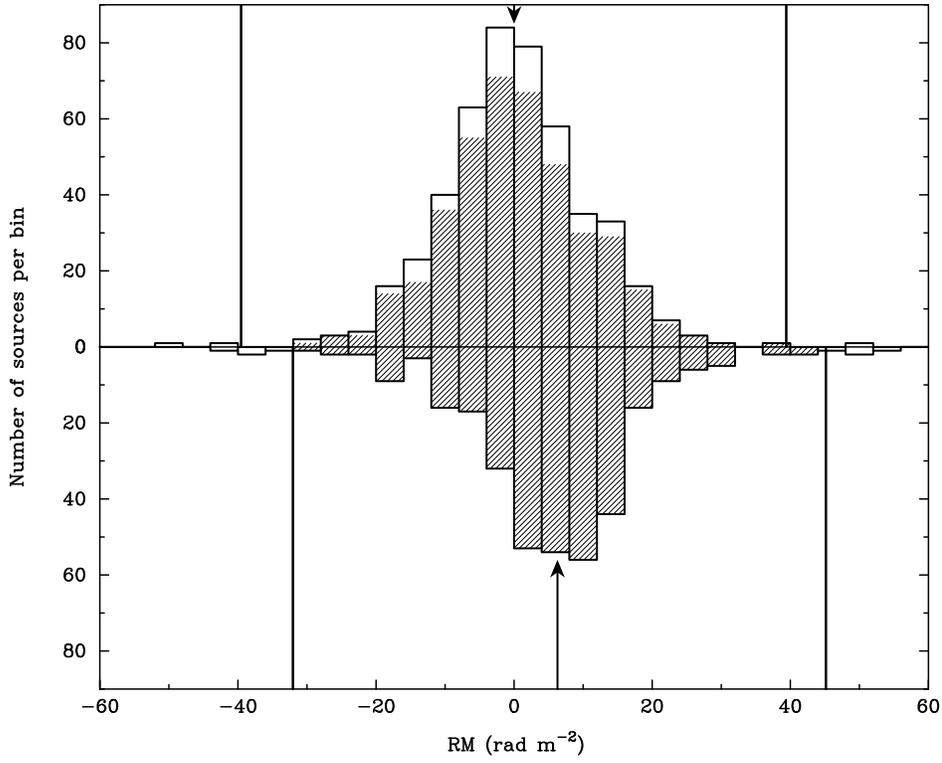}}
\caption{Histogram of the RM distribution towards the north Galactic pole in the top panel and the south Galactic pole in the bottom panel, binned every 4 rad m$^{-2}$. The histograms include all raw RMs (solid outlined histograms) as well as the final RMs after removing extrema and anomalous regions (shaded histograms). The thick vertical lines indicate the 3.2 $\sigma$ cut-off boundary for the extreme RM rejection scheme. The arrow in the top panel indicates the median (0.0 rad m$^{-2}$) of the final north Galactic cap RM data set, while the arrow in the bottom panel indicates the median ($+$6.3 rad m$^{-2}$) of the final south Galactic cap RM data set. The standard deviations of the final data sets are 9.2 rad m$^{-2}$ and  8.8 rad m$^{-2}$ towards the north and the south Galactic pole respectively. }
\label{fig:histo}
\end{figure}

\begin{figure}
\centering
\includegraphics[width=0.6\textwidth]{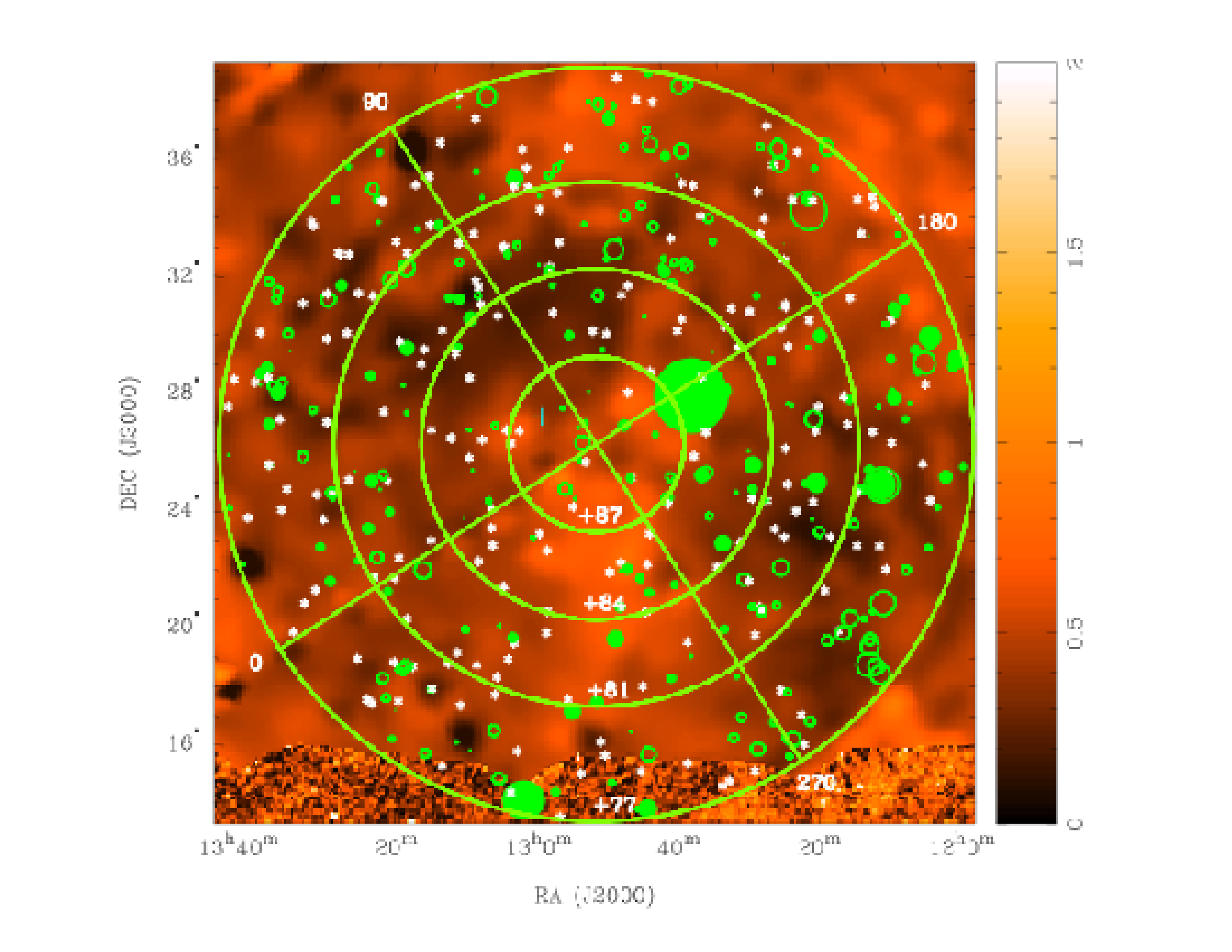}
\vspace{.05in}
\includegraphics[width=0.6\textwidth]{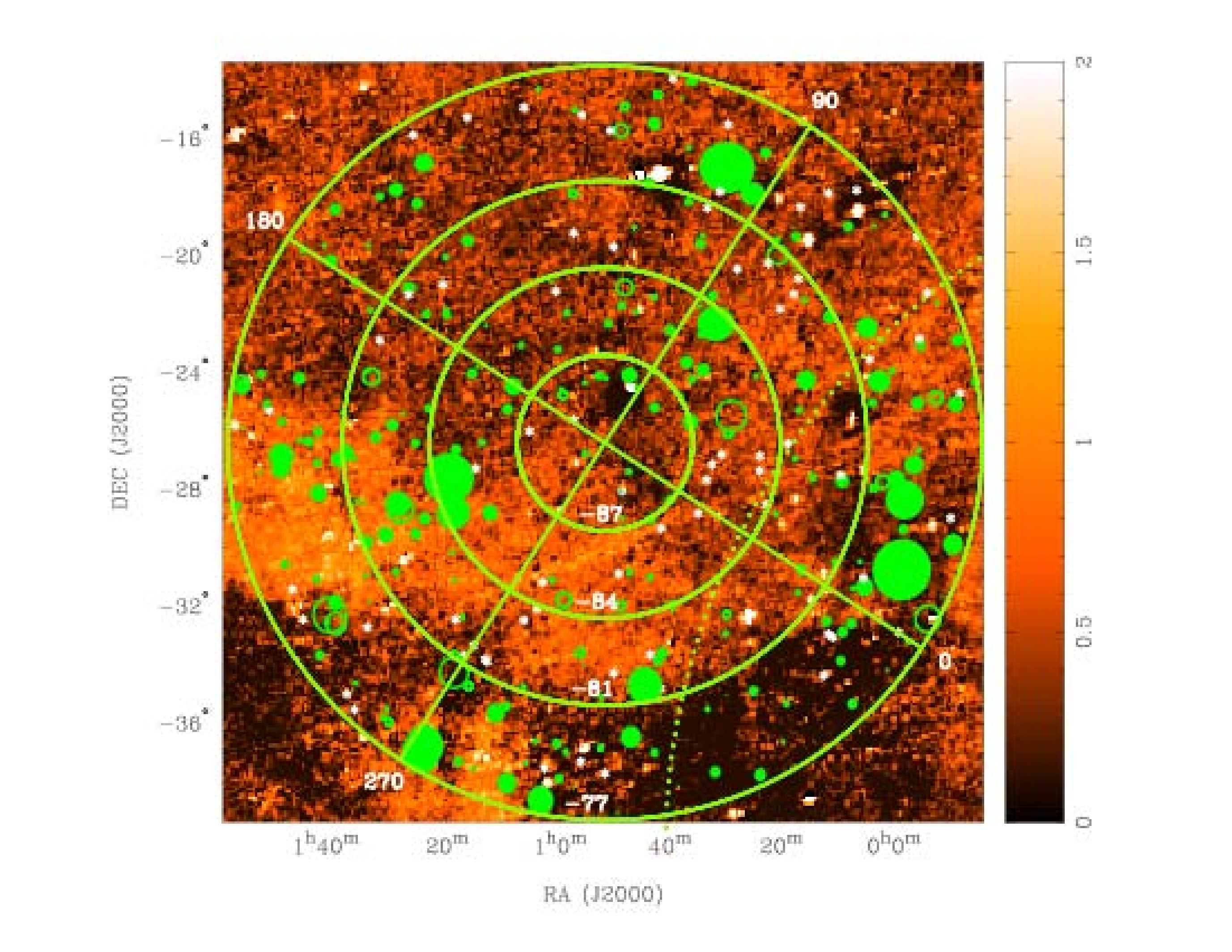}
\epsscale{1}
  \caption{\footnotesize Rotation measure distribution overlaid on the H$\alpha$ emission map \citep{finkbeiner2003} towards the north Galactic pole in the top panel and towards the south Galactic pole in the bottom panel (after filtering sources with low signal-to-noise ratio or high polarized fractions; sources with extreme value of RMs or in anomalous RM regions are still shown). The color scale is in units of Rayleighs.  Positive (Negative) RMs are denoted by filled (open) circles with their diameters proportional to the magnitude of RM. The largest circle corresponds to a RM value of +93 (+76) rad m$^{-2}$ towards the north (south) Galactic pole. Sources with RMs consistent with zero at 1 $\sigma$ are denoted by asterisks. The blue cross in the top panel indicates the centre of the Coma cluster. The dotted green curve in the bottom panel marks the projection of the LB wall: sight lines to the east of the boundary penetrate the LB wall. The resolution change at the bottom of the top image corresponds to the change of H$\alpha$ data being used to create the composite map from the WHAM to the Southern H-Alpha Sky Survey Atlas (SHASSA), the latter has better spatial resolution but much poorer sensitivity.  The H$\alpha$ image towards the south Galactic pole is a combination of  SHASSA and WHAM data in most of Galactic Quadrant 2 and 3 and it consists of SHASSA data only in Quadrant 1 and 4.}
  \label{fig:rm_dis}
\end{figure}

The RM derivation procedures described in the previous section produced 472 reliable RMs towards the north Galactic pole (94 \% of the total number of observed sources) and 341 (72 \% of the total number of observed sources) towards the south. The source coordinates, RMs, RM uncertainties, and their flux information are listed in Table~\ref{table:northrms} and~\ref{table:southrms} for sources towards the north and the south cap respectively. To help visualizing the distribution of RMs towards the Galactic caps, we present the histograms of the unfiltered RMs in Fig~\ref{fig:histo}. They are shown as solid outlined histograms which appear to be Gaussian with few outliers.  The spatial RM distribution towards the north and south Galactic caps are presented in the top and bottom panel of Fig~\ref{fig:rm_dis} respectively. 
In this section, we will address how outliers and anomalous RM regions were identified. We do so to ensure that the RMs towards the Galactic caps are representative of the large scale Milky Way foreground with minimal contributions from elsewhere along the sight line. 

\subsection{Features towards the Galactic poles at other wavelengths}
\label{subsection:NPS}
We have visually inspected multi-wavelength images towards the Galactic caps to look for possible features seen at other wavelengths that might be associated with RM structures. In particular, we have examined the following maps: the HI neutral hydrogen map \citep{kalberla2005}; the dust map \citep{schlegel1998}; the ROSAT all sky survey X-ray map \citep{snowden1997}; the 408 MHz total synchrotron intensity map \citep{haslam1982}; the 1.4 GHz polarized synchrotron emission map (north: \citealt{wolleben2005}, south: \citealt{testori2008}), and the H$\alpha$ all sky composite map \citep{finkbeiner2003}. 

We note that the Coma cluster (the brightest X-ray feature in the ROSAT all sky X-ray image towards the north Galactic cap) is located only 2$^\circ$ away from the Galactic pole. According to \cite{kim1990}, the Coma cluster has a magnetic field strength of $\sim$ 2 $\mu$G and it can produce an excess RM of several 10s of rad m$^{-2}$. Hence, we have excluded 10 EGSs within 2$^\circ$ of the Coma cluster (RA=12:59:48.7, DEC=+27:58:50.0) to ensure that our RM sample is free of contributions from the intragalactic medium of Coma. These rejected EGSs are denoted by superscript a in column 8 of Table~\ref{table:northrms}.

Towards the north Galactic pole, part of  the North Polar Spur (NPS), a bright polarized feature at radio wavelengths and a prominent X-ray feature at high positive Galactic latitude, is in our field of view. As pointed out by \cite{frick2001}, the NPS can produce a maximum foreground RM of $\sim$ $-$28 rad m$^{-2}$ towards $\ell$=289$^\circ$, $b$=+23$^\circ$, which corresponds to an internal magnetic field strength of $\sim$ 0.9 $\mu$G \citep{frick2001,heiles1980}. This internal magnetic field strength of the NPS coupled by thermal electrons can produce significant RMs in our field of view . The polarized emission of the NPS and its counterpart below the Galactic plane in the DRAO 1.4 GHz polarization survey has been modeled by  \cite{wolleben2007} with two overlapping spherical synchrotron emitting magnetized shells where the Sun resides in one of them. The model correctly predicts the morphology of the NPS polarized emission with the projection of the overlapped region onto the sky roughly coincides with the NPS X-ray emission. Although the two-shell model does not include predictions of RMs through the shells,  one expect complicated RM structure for sight lines that pass through both shells. Therefore, we choose to discard all EGSs whose sight line penetrate the polarized radio feature and the prominent X-ray feature of the NPS ($b$$<$+80$^\circ$ and 50$^\circ$$<$$\ell$$<$310$^\circ$). A total of 56 EGSs are rejected and they are indicated by superscript b in column 8 of Table~\ref{table:northrms}. At high negative Galactic latitude, there are no features seen at other wavelengths that might contaminate RMs.

\begin{figure}
\centering
\includegraphics[width=0.6\textwidth]{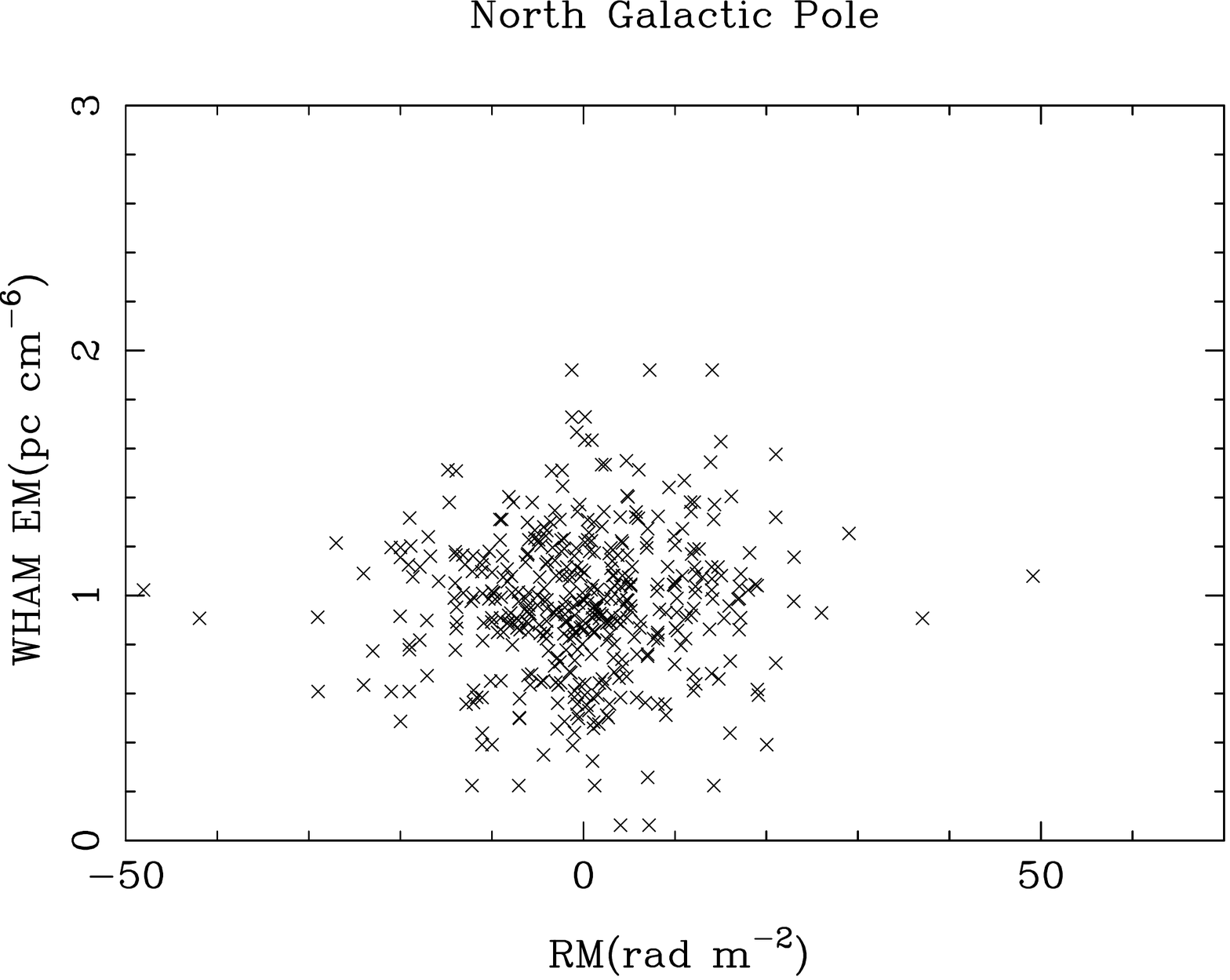}
\vspace{.05in}
\includegraphics[width=0.6\textwidth]{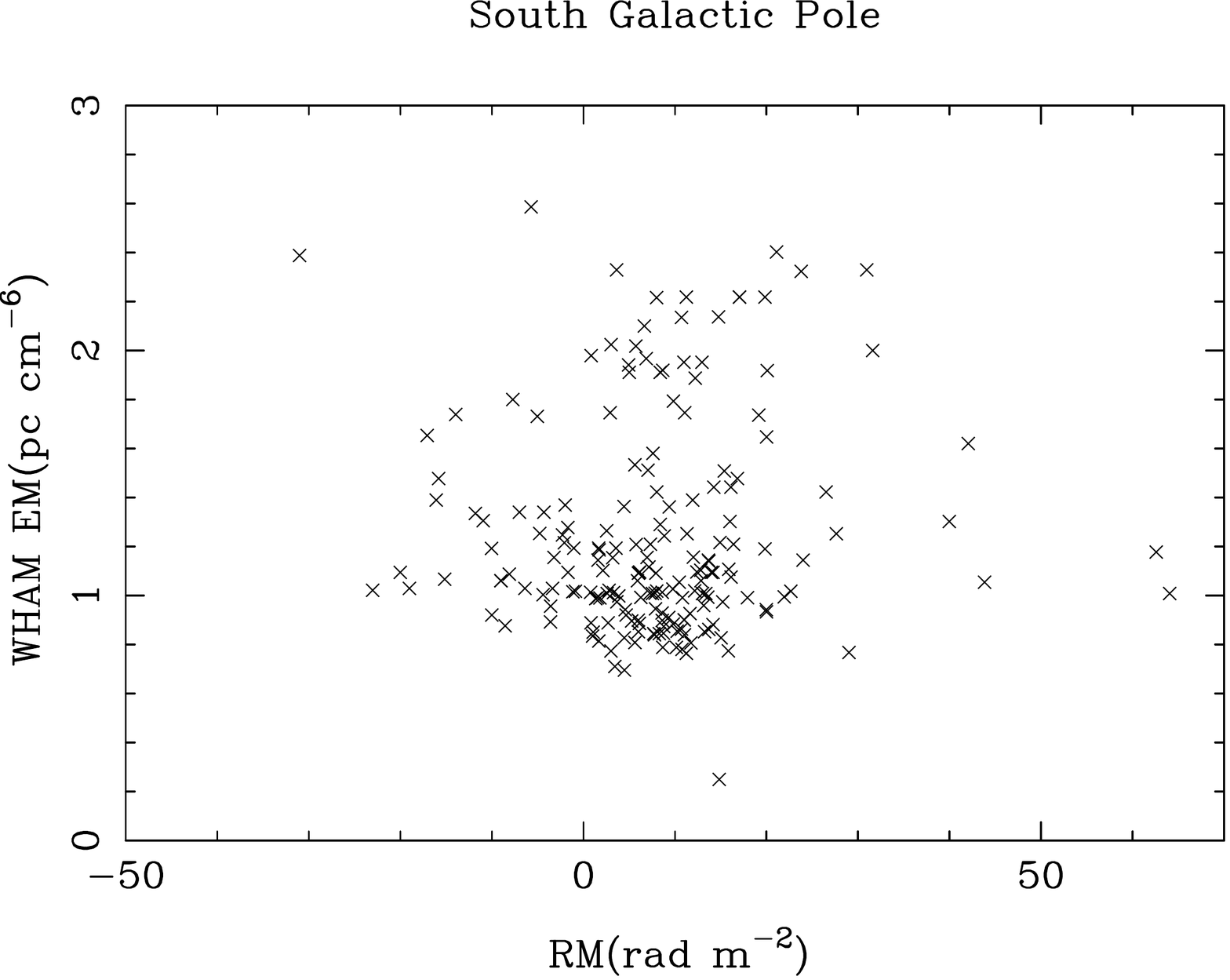}
\epsscale{1}

\caption{The scatter plot of EMs (from WHAM) against RMs along sight lines to EGSs towards the north (top panel) and the south (bottom panel) Galactic caps. All EGS sight lines towards the north Galactic pole have corresponding WHAM EM measurements, while only 192 sight lines towards the south Galactic pole has corresponding WHAM EM measurements. For clarity, measurement errors associated with EM and RM are not plotted. No clear trend of correlations between EM and RM can be seen.}
\label{fig:em_rm}
\end{figure}

One might expect some correlation between RM and H$\alpha$ as they both depend on the electron density content along a sight line. Fig~\ref{fig:rm_dis} shows the distribution of RMs towards the Galactic cap regions overlaid on the \cite{finkbeiner2003} H$\alpha$ all sky composite map. No evidence of correlations between RM and EM could be seen. We have further investigated this issue by making a scatter plot of EM versus RM towards the north and the south Galactic cap (Fig~\ref{fig:em_rm}). One can clearly see that points in these plots appear to be randomly located without any trend. The linear correlation coefficient between RM and $\sqrt{\rm EM}$ is $\sim$ 0.04 towards the north Galactic cap and $\sim$ 0.03  towards the south Galactic cap hence there is little support for correlations between EM and RM towards the Galactic poles. It is not surprising, however, to find RM and EM uncorrelated. As EM is proportional to the electron density squared, it is more sensitive to high density regions; while RM is proportional to just the electron density, thus it  has more contributions from diffuse low density regions that occupy greater line-of-sight path lengths. Magnetic field reversals along the line of sight can also remove any possible correlations between RM and EM.  All these effects work towards decoupling the measured RM and EM along a particular sight line. 

\subsection{Outliers}
\label{subsection:outlier}
Besides excluding features at other wavelengths that might contribute to the observed RMs, one also has to remove sources in the sample with high intrinsic RMs. Typical radio galaxies at a few GHz are generally dominated by their bright radio lobes which are thought to have small intrinsic Faraday rotation measures \citep[see for example,][]{willis1978,feain2009}. Since we have selected EGSs based only on their polarized flux and $|$$b$$|$, our sample inevitably contains core-dominated EGSs which have high internal RMs (possibly generated from a cocoon filled with magnetic fields and thermal electrons) and EGSs which lie along line of sights to multiple intervening galaxies or clusters with high RMs by chance. We have employed the Chauvenet's criterion\footnotemark[8]\footnotetext[8]{The Chauvenet's criterion states that a data point can be discarded with reasonable confidence if less than half an event is expected to be farther from the median than the suspect point \citep{bevington2003}.}. For a sample size of $\sim$ 400, one expects less than 0.5 source with an RM that deviates more than 3.2 $\sigma$ from the median. EGSs with RM values  3.2 $\sigma$ away from the median of the distribution are discarded\footnotemark[9]\footnotetext[9]{The Chauvenet's criterion is applied to the RM data set towards the north cap after regions coinciding with the Coma cluster and the NPS have been removed.}. This is based on the assumption that the underlying RM distribution is Gaussian. The thick vertical lines in Fig~\ref{fig:histo} indicate boundaries beyond which RMs have been rejected. These EGSs are indicated in column 8 of Table~\ref{table:northrms} and~\ref{table:southrms}. 

\subsection{Smooth RM maps and anomalous RM regions}
\label{subsection:badregion}
\begin{figure}
\centering
\includegraphics[width=0.6\textwidth]{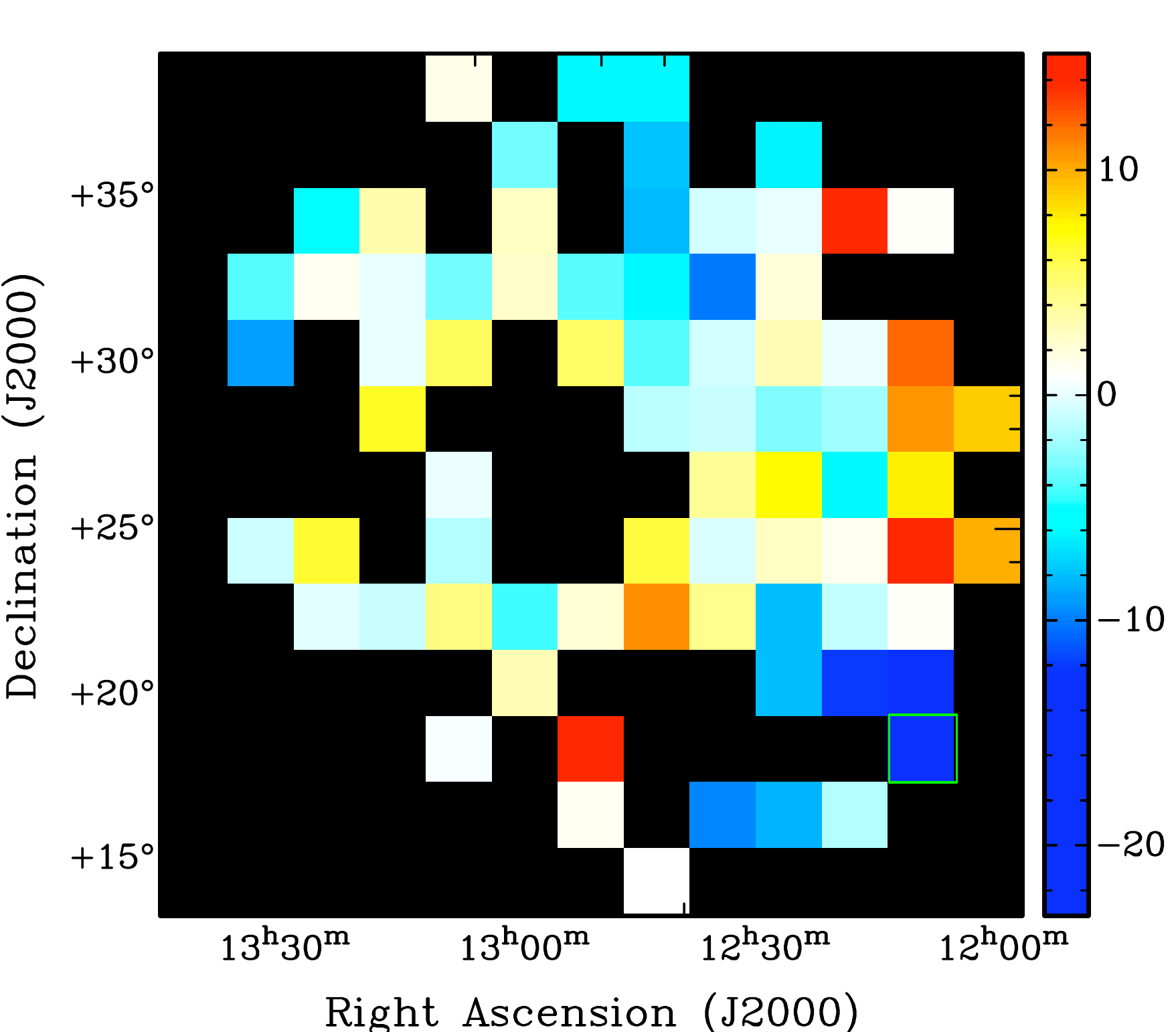}
\vspace{.05in}
\includegraphics[width=0.6\textwidth]{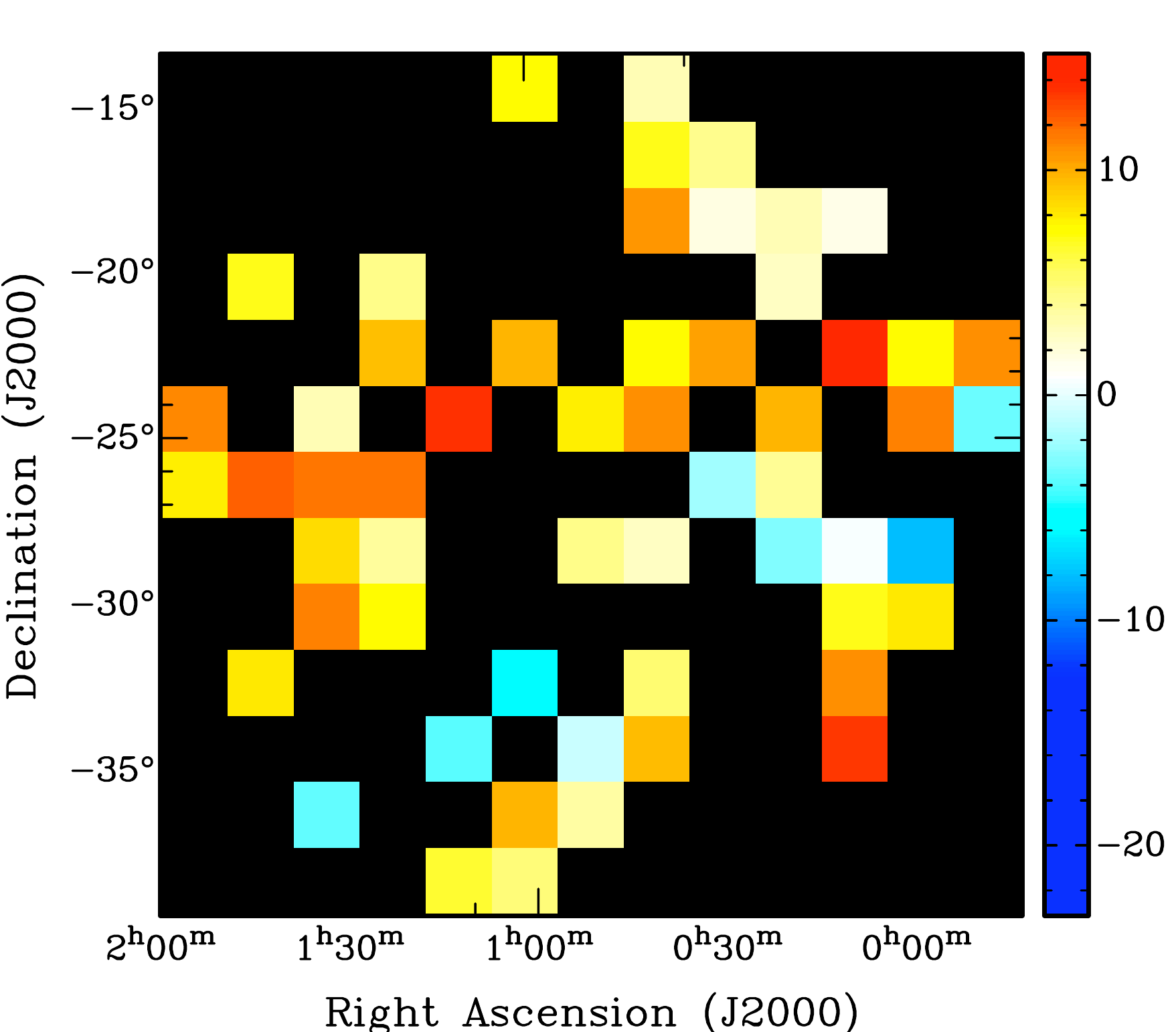}
\epsscale{1}
  \caption{\footnotesize The smooth RM map towards the north (top panel) and the south (bottom panel) Galactic cap in SIN projection. The Galactic poles are located at the center of the images. The median RMs in 2$^\circ$ by 2$^\circ$ pixels are computed. Blanked regions represent pixels with insufficient number of EGSs ($<$3 ) to obtain reliable statistics. The color scale to the right of the figure is in units of rad m$^{-2}$ and is chosen such that 0 rad m$^{-2}$ is represented by white. Pixels values that deviate more than 1.64 $\sigma$ from the median of the entire RM distribution are regarded as anomalous regions and are indicated by green rectangles. One such cell is identified towards the north Galactic cap and consequently, the 3 RMs within the cell are discarded for the purpose of studying the large scale Galactic magnetic field. No anomalous RM region is identified towards the south Galactic cap.}
\label{fig:smooth_rm}
\end{figure}

After discarding EGSs in the cap regions which have large intrinsic RMs, subregions towards the Galactic caps where RMs differ systematically from the overall smooth behavior expected from a large scale magnetic field should also be identified. To locate these subregions, we have first constructed smooth RM maps towards the Galactic caps using the region-filtered (without the NPS and Coma Cluster regions) and outlier-free RM sample. We have divided the cap regions into  2$^\circ$ $\times$ 2$^\circ$ cells and computed the median RM of the EGSs within the individual cells. To ensure the statistics are reliable, we require at least 3 or more EGSs in a cell to derive a valid median RM. The resulting smoothed maps are shown in  Fig~\ref{fig:smooth_rm}.
One can see clearly from the smoothed maps that the RM towards the north Galactic pole is distributed randomly around zero whereas the overall RM towards the south Galactic pole is positive. Next, we have calculated how many standard deviations the median of RMs in a cell deviates from the median of the entire RM sample. We discard all sources within a cell if the cell has a minimum of 3 EGSs and has a median RM that deviates more than 1.65 $\sigma$ ($>$ 90\%) away from the overall median of the RM distribution. We have identify 1 such cell towards the north Galactic cap and it is indicated by green rectangle in the top panel of Fig~\ref{fig:smooth_rm}. No anomalous RM cell  was identified towards the south Galactic cap as shown in the bottom panel of Fig~\ref{fig:smooth_rm}. We are aware that this rejection criterion is only sensitive to anomalous RM regions with scales comparable to the 2$^\circ$ $\times$ 2$^\circ$ cells and that the boundary of the cells is chosen arbitrarily. We have verified that sources being rejected by this algorithm are very similar regardless of the choice of the cell size. This leads us to believe that the method outlined above is a robust way of rejecting anomalous RM regions. The 3 sources towards the north Galactic cap discarded by applying the above criterion are indicated in column 8 of Table ~\ref{table:northrms}. After the rejection procedures described above, there remains 400 RMs towards the north Galactic cap and 329 RMs towards the south Galactic cap. These RMs should well represent the large scale magnetic field towards the Galactic poles.

\subsection{Rotation Measure structure function}
\label{subsection:rmsf}
To study the RM structures towards the Galactic caps and to understand the properties of turbulence at high $|b|$, we have computed the second-order 1D RM structure function (SF) defined as 
\begin{equation}
\label{eq:rmsf}
SF_{\rm RM,obs}(r)=\langle [RM(x)-RM(x+r) ]^2 \rangle, 
\end{equation}
where $r$ is the projected angular separation between a pair of extragalactic sources and angular brackets denote the expectation value which involves taking the ensemble average of independent measurements with the same range in angular separation $r$. The uncertainties of RM measurements ($\sigma_{\rm RM}$)  contribute to the observed RM structure function in the form of a DC offset \citep[see Appendix A of][]{haverkorn2004}. The offset-corrected structure function is given by
\begin{equation}
\label{eq:rmsf2}
SF_{\rm RM} (r) = SF_{\rm RM,obs} (r)- SF_{\sigma_{\rm RM}} (r). 
\end{equation}

Since RMs of different components of the same EGS will contribute many square differences but they do not provide independent information \citep{minter1996}, we have identified EGSs in the filtered RM data sets that appear to be associated with each other using the NASA/IPAC Extragalactic Database (NED) and excluded them from the structure function calculations. We defer the study of RM fluctuation of these closely spaced EGSs to a forthcoming paper. The RM structure functions were constructed using 354 RMs towards the north and  319 RMs towards the south Galactic cap. This is the largest and most reliable RM data set available to date to study turbulence at high Galactic latitudes -- the high source density allows us to probe a large range of scales from $\sim$ 0.1$^\circ$ up to 26 $^\circ$, while the large number of RMs allows one to determine the SF accurately based on 62,481 correlations towards the north and  50,721 towards the south Galactic poles.

\begin{figure}
\centering
\includegraphics[width=0.6\textwidth]{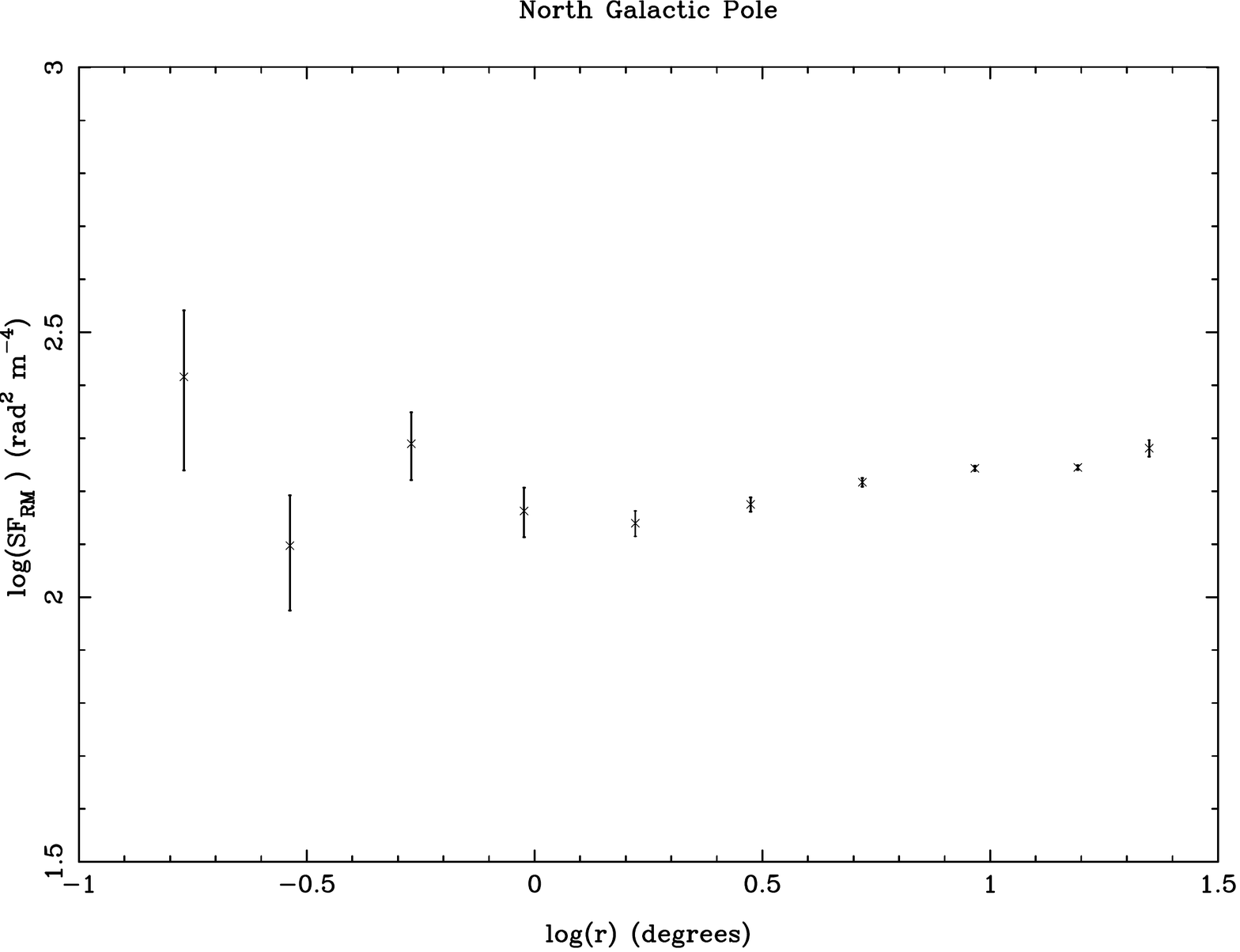}
\vspace{.05in}
\includegraphics[width=0.6\textwidth]{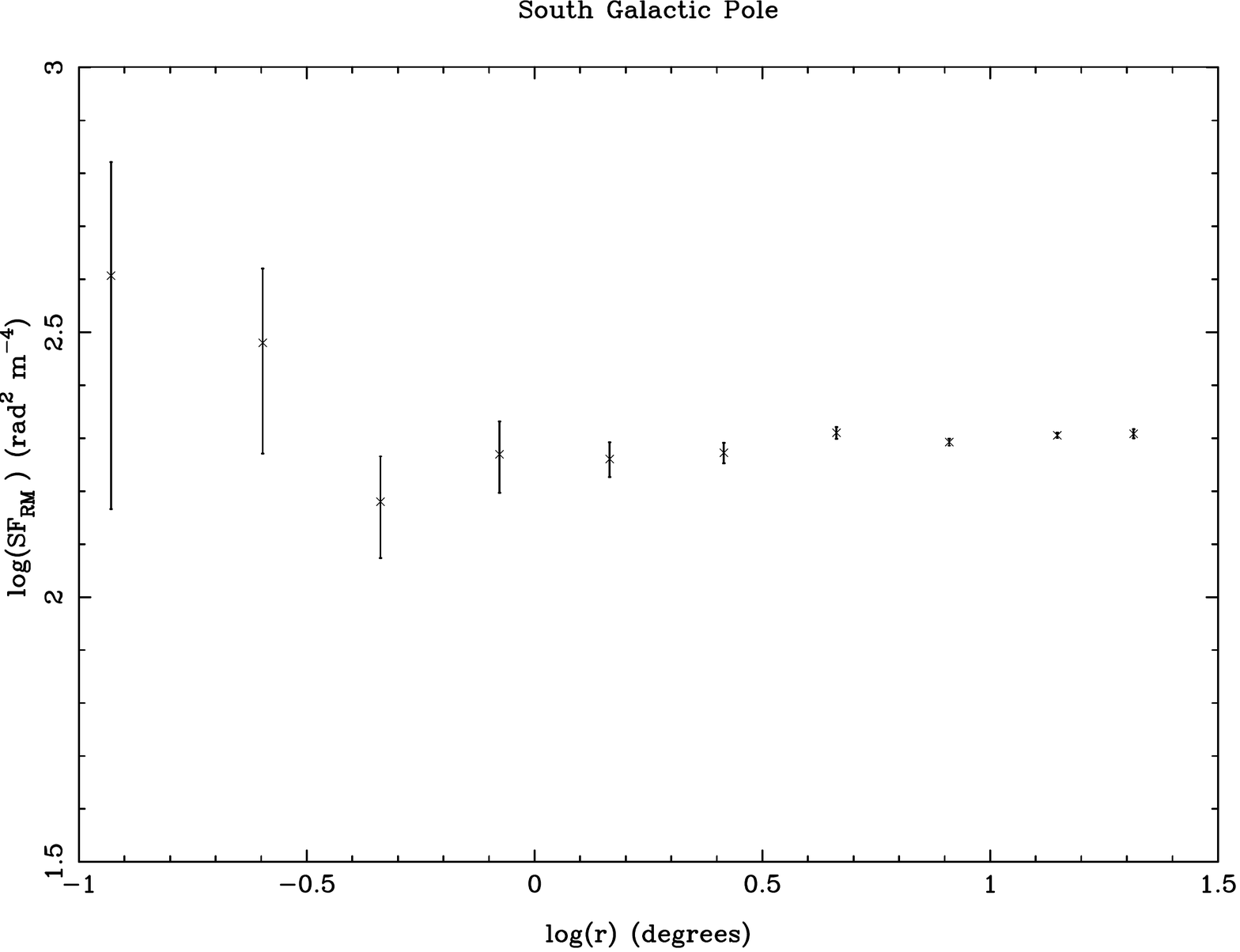}
\epsscale{1}
\caption{The RM structure function towards the north (top) and the south (bottom) Galactic cap computed using Eq ~\ref{eq:rmsf} and ~\ref{eq:rmsf2}. The structure function has been binned in equal log interval (0.25) and the minimum number of correlations per bin is 10. The error bars denote the standard error in the mean of each bin. }
\label{fig:rmsf}
\end{figure}

The resulting RM SF towards the north and the south Galactic caps are plotted in the top and bottom panel of  Fig~\ref{fig:rmsf} respectively. A power-law fit is performed on the observed RM SFs - the slope of the SF towards the north Galactic cap is $+$0.08 $\pm$ 0.01 while that towards the south is $+$0.03 $\pm$ 0.01. Since the structure function measures the fluctuation of RM along different sight lines, it is not sensitive to  constant RM contributions in the cap regions, such as that from a constant vertical magnetic field. However, the RM SF is sensitive to any large scale magnetic field or electron density gradients in the field of view. For example, a large scale horizontal magnetic field towards the Galactic poles would produce more structures on the largest scales and would lead to a rising structure function. The nearly flat structure functions towards both caps confirm our visual inspection of Fig~\ref{fig:smooth_rm} that RMs are uncorrelated and that there is no preferred scale of RM structure towards the Galactic caps.  The flat RM SFs towards the Galactic poles are consistent with the non-detection of horizontal fields towards the poles in \S~\ref{subsection:horizontal}.

\subsection{The RM distribution towards the Galactic poles}
\label{subsection:rmdist}
The distribution of the final RM data sets is shown in the shaded histograms in Fig~\ref{fig:histo}. The median RM towards the north Galactic pole is 0.0 $\pm$ 0.5 rad m$^{-2}$, which indicates that the median RM towards the north Galactic cap is consistent with zero. The 3 $\sigma$ upper limit on the $|$RM$|$ towards the north Galactic pole is 1.4 rad m$^{-2}$. On the other hand, the median of RMs towards the south Galactic cap is $+$6.3 $\pm$ 0.7 rad m$^{-2}$. The RM distribution towards the south Galactic cap is inconsistent with zero at greater than 9 $\sigma$ level.

\subsection{Comparison with \cite{taylor2009} }
\label{subsection:rmcompare}
\begin{figure}
\centering
\includegraphics[width=0.6\textwidth]{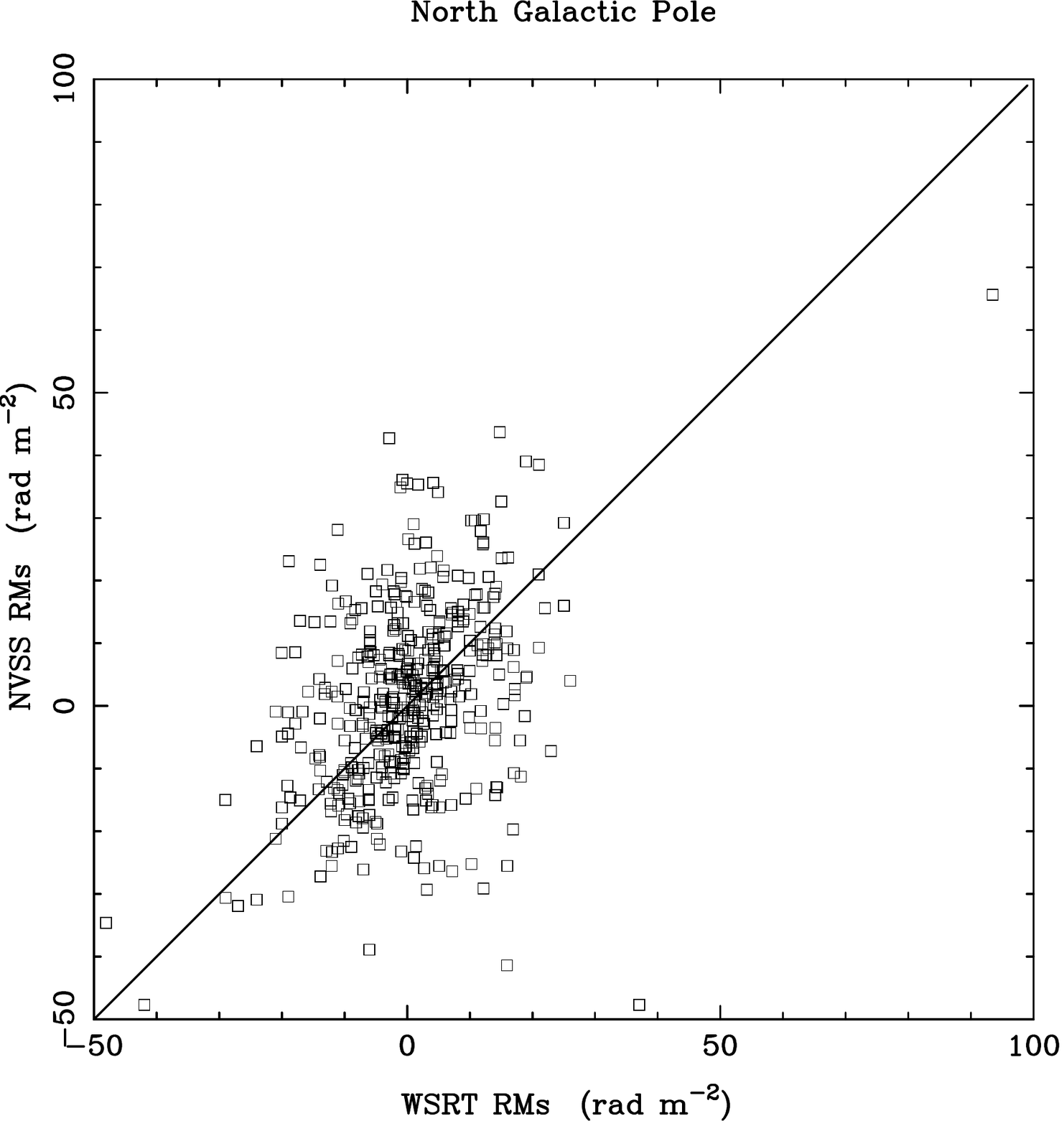}
\vspace{.05in}
\includegraphics[width=0.6\textwidth]{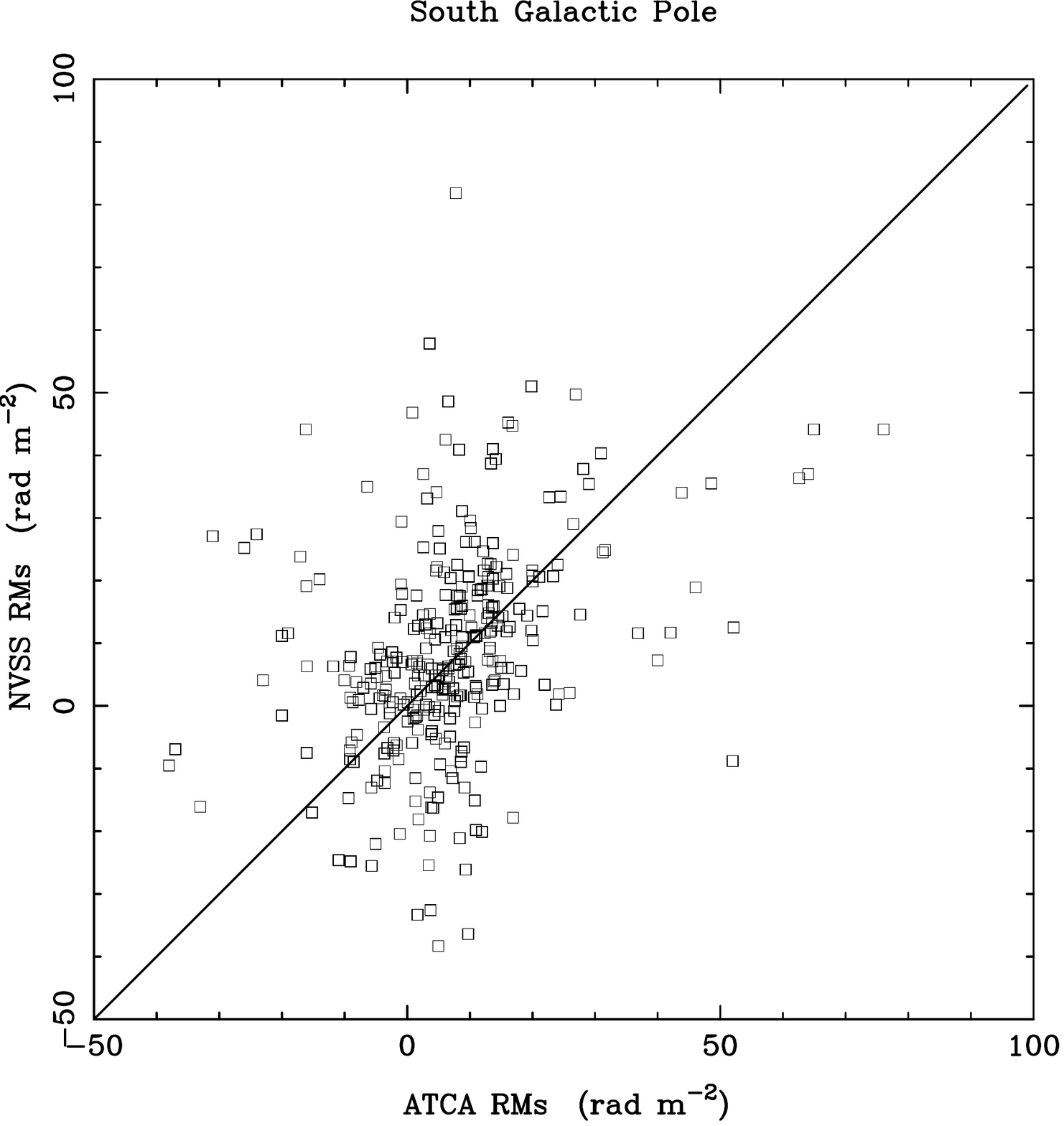}
\epsscale{1}
\caption{The comparison between NVSS RMs derived by \cite{taylor2009} and our RMs. NVSS RMs are plotted against WSRT (ATCA) RMs towards the north (south) Galactic pole in the upper (lower) panel. The solid line indicates where NVSS RMs and our RMs agree with each other. For clarity, error bars associated with the measurements are not drawn.}
\label{fig:rmcompare}
\end{figure}

As mentioned in \S~\ref{section:introduction}, \cite{taylor2009} have computed RMs from the NVSS catalogue. In Figure~\ref{fig:rmcompare}, we have plotted the RMs derived by \cite{taylor2009} against the RMs that we have derived for the same sources. The thick solid line of slope 1 in the figure indicates where sources should lie if the NVSS RMs and our RMs are equal. Approximately  57\% of the RMs from the two samples towards the north cap agree with  each other within their measurement errors, whereas this percentage is 53\% towards the south cap\footnotemark[10]\footnotetext[10]{One would expect approximately 68\% of the RMs from the two studies to be consistent within 1 $\sigma$ if there were no systematic errors.}. We do not find the $\pm$ 650 rad m$^{-2}$ RM ambiguity problem as discussed by \cite{taylor2009} in our high latitude data set because the measured RMs have small magnitudes. The linear correlation coefficients between the two RM data sets are:  0.39 towards the north and 0.36 towards the south, indicating that there exists substantial difference between the two RM samples. The signs of the  \cite{taylor2009} RMs and our RMs do not always agree - this is demonstrated by sources located in the upper left and lower right corners of the scatter plots. The difference between \cite{taylor2009} RMs and our WSRT/ ATCA RMs is most likely due to multiple RM components that produces non-linearity in polarization angle against $\lambda$$^2$ relation that  \cite{taylor2009} was not able to identify, and potentially ionospheric RM which depends on the observing time, site and also epoch in the solar cycle. While RMs derived in \cite{taylor2009} can be used collectively to describe the large scale Galactic magnetic field by averaging over large areas, extra care must be taken if one plans to use the individual RM values in their catalogue since these RMs can potentially be inaccurate. Our RM catalogue is more suitable if one plans to investigate the RM properties of individual sources.

\section{The origin of the observed high Galactic latitude RM structure}
\label{section:discussion}
In this section, we explore possible origins of the observed RM pattern towards the Galactic caps: an overall RM consistent with zero towards the north and a positive median RM towards the south. We compare predictions from different models with the observed RM pattern. We first investigate local sources/ events that might give rise to the observed RMs: the local interstellar medium, the local bubble and a Parker instability loop\footnotemark[11]\footnotetext[11]{We have excluded the NPS two shell model \citep{wolleben2005} from consideration as we have already discarded RMs of EGSs whose sight lines intercept the NPS (see \S~\ref{subsection:NPS})}. We then estimate the magnetic field strength towards the Galactic poles from the observed RMs by assuming that the observed Faraday rotation occurs in diffuse ionized gas in the Galaxy. Finally, we consider the likelihood that the observed RM pattern has been generated by global events in the Galaxy, such as a large scale Galactic wind, large scale Galactic dynamos or a relic field.

\subsection{Local origins}
\label{subsection:local}
\subsubsection{Local Interstellar medium}
\label{subsection:lism}

The immediate medium surrounding the sun could potentially produce the observed RM pattern at high Galactic latitude. Even though the sun resides in a low density cavity called the Local Bubble (LB) (see more in \S~\ref{subsection:localbubble} ), there are warm partially ionized clouds surrounding the Sun \citep{linsky2007}.
These clouds are magnetized and have free electrons, and therefore are capable of rotating the polarization plane of incident radiation. \cite{spangler2009} estimated the upper limit of $|$RM$|$ produced by the LISM to be 0.32$-$1.1 rad m$^{-2}$ by assuming that typical clouds have an electron density of 0.12 cm$^{-3}$, a volume filling factor of 5.5$\%$ - 19$\%$  \citep{redfield2008a,redfield2008b} and a magnetic field strength of 4 $\mu$G. Different studies have yield different estimates for the local magnetic field strength.  \cite{snowden1998} suggested that a magnetic field of up to 7$\mu$G is required in the LB to counter balance the enormous thermal pressure ($p/k$$\sim$ 15,000 cm$^{-3}$ K ) exerted by the enclosing hot X-ray gas, but the recent discovery of X-ray emission associated with charge exchange between solar wind ions and heliospheric plasma has greatly alleviated the need of non-thermal pressure support of the LB and has lowered the required magnetic field strength to $\sim$ 2.8 $\mu$G \citep{welsh2009a}. Other works \citep{opher2007,wood2007} find a local field strength of  $\sim$ 2 $\mu$G. Since various estimations of the local magnetic field strength are lower than the 4 $\mu$G adopted by \cite{spangler2009}, the LISM $|$RM$|$ can be a factor of two smaller than the estimates of \cite{spangler2009}. We note that the estimated LISM $|$RM$|$ contribution of $\sim$ 1 rad m$^{-2}$ is an upper limit as $|$RM$|$ would be much smaller if the magnetic field reverses within the clouds or from cloud to cloud. As the typical RM measurement error in our experiment is a few rad m$^{-2}$, contribution of RM from the LISM is likely to be negligible.

\subsubsection{Effects of the local bubble wall}
\label{subsection:localbubble}

In this section, we consider the possibility that the Faraday rotation originates in the wall of the Local Bubble. As mentioned in \S~\ref{subsection:lism}, the Sun is situated in a low-density cavity thought to be created by star formation and subsequent supernova explosions that occurred in the past 25-60 Myrs \citep{frisch2007}. Such processes would sweep up materials and magnetic fields in the solar neighborhood into a dense shell and might produce measurable RM for sight lines through it. 

To test if the LB wall is responsible for producing the observed RM towards the Galactic caps, we have adopted the LB wall model constructed by \cite{cl2002,cl2003} with slight modifications: we have modeled the LB wall as a slanted cylinder of constant radius 0.085 kpc that extends to 0.2 kpc both above and below the Galactic plane. Unlike the original model in \cite{cl2002,cl2003}, our cylindrical wall has open ends since the recent expansion of NaI measurements made by \cite{lallement2003} and \cite{welsh2009b} find no continuous neutral LB boundary at high Galactic latitudes. We have computed the projection of the modeled LB boundary towards the Galactic poles and plotted it in Fig~\ref{fig:rm_dis} as a green dotted line. None of the sight lines towards the north Galactic cap intercepts the local bubble wall, while 267 out of 341 sight lines intercept the modeled wall towards the south Galactic cap. Towards the south Galactic pole, the median RM of sight lines that penetrate the LB wall is $+$6.1 $\pm$ 0.7  rad m$^{-2}$ whereas those do not penetrate the LB wall have a median of $+$ 7.8 $\pm$ 1.8 rad m$^{-2}$. Since the inferred RM through the LB wall is consistent with zero ($-$2 $\pm$ 2 rad m$^{-2}$), we conclude that the local bubble wall is unlikely to be the major contributor to the observed RM towards the caps if the adopted wall model is realistic. 

The  3$\sigma$ upper limit of $|$RM$|$ $\sim$  7 rad m$^{-2}$ through the local bubble wall can be used to infer the magnetic field strength in the LB wall. Since \cite{bhat1998} found that the scintillation measures of 20 nearby pulsars were well modeled by a scattering structure with local electron density enhancement  of a factor of 10 which roughly coincides with the neutral LB wall, we assume that the electron density wall can be well traced by the neutral wall. If energetic events have swept up magnetic fields and electrons of initial density 0.025 cm$^{-3}$ \citep{bhat1998} into a cylindrical shell of radius $R$ $\sim$ 85 pc \citep{cl2002,cl2003} centered at the Sun, the thickness of the shell in pc ($\delta R$) is related to the electron density enhancement ($x$$=$10) in the shell and its radius $R$ by conserving the mass of electrons before and after the formation of the shell by $\delta R = \frac {R}{2x}$. The thickness of the cylindrical LB shell is estimated to be $\sim$ 4 pc. If the RM produced by the wall is  7 rad m$^{-2}$, it implies a magnetic field strength of $\sim$ 9 $\mu$G . This prediction from a simple theoretical model of the LB is in rough agreement with the magnetic field strength of  8 $\mu$G estimated by \cite{andersson2006} \citep[or earlier][]{leroy1999}, who have applied the \cite{chandrasekharfermi1953} method on starlight polarization measurements  towards stars at distances from 40$-$200 pc in the direction of $\ell$=300$^\circ$, $b$=0$^\circ$. 

\subsubsection{Small Scale outflows: a Parker instability loop}

\cite{parker1979} demonstrated that a system of horizontal magnetic field and cosmic rays in a vertical gravitational field can be unstable with respect to the bending of magnetic field lines in the vertical plane. For example, energetic stellar events can produce such waviness in a Galactic disk \citep{kronberg1994}. It is possible that the entire surface of the Galactic disk is packed with Parker loops with height $\sim$ 1 kpc and width of 0.1$-$1 kpc \citep{parker1992}. If these magnetic loops thread the warm ionized gas in the Galaxy, it is plausible that they contribute towards the observed RM patterns at high Galactic latitude.

\cite{frick2001} performed a wavelet analysis on the all sky RM distribution and found that the RM structure at the largest scale has been shifted to a negative Galactic latitude of $-$15$^\circ$, even after omitting sight lines that intercept Loop I. While one can interpret this shift as due to a stronger large-scale field in the southern Galactic hemisphere, possibly due to a separate halo dynamo with opposite parity to the disk field, it can also be due to a Parker instability loop. The authors suggest a scenario that a Parker instability loop with the Sun located near its top ($\sim$ 50 pc above the galactic plane), a horizontal extent of $\sim$ 400 pc and a magnetic field strength enhancement of 0.5 $\mu$G can shift the symmetry axis of the RM structure to $b$=$-$15$^\circ$. While the physical parameters of the inferred loop is consistent with that proposed by \cite{parker1992}, it has difficulties explaining the observed RM pattern towards the Galactic caps on its own. Since the sun is located near the top of this loop, when looking towards a cone of radius 13$^\circ$ around the Galactic poles, the extra $|$RM$|$ produced by the loop is small $\sim$  1-2 rad m$^{-2}$, comparable to individual RM measurement errors. Furthermore, the median RM towards the south Galactic cap would be zero if the loop is sufficiently symmetric  as half of the south cap region should have the exact opposite RMs to the other half. This is inconsistent with what has been observed: the RM towards the entire south Galactic cap is positive (Fig~\ref{fig:smooth_rm}). If the Sun is not located exactly at the top of this loop, then one would expect median RMs of different signs towards each cap because the magnetic field lines should be continuous. This is again inconsistent with the observations because the average RM towards the north cap is consistent with zero. We therefore conclude that the Parker instability loop proposed by \cite{frick2001} alone cannot reproduce our high latitude RM measurements.

We note that the argument above is based on the assumption that Parker instability loops exist in the warm ionized gas (typical WIM density $\sim$ 0.1  cm$^{-3}$) in the Galaxy. If these instabilities are associated with the hot phase of the ISM where the typical magnetic field strength is $\sim$ 0.1 $\mu$G  \citep{beck1996} and the typical density is very low 10$^{-3}$ - 10$^{-4}$ cm$^{-3}$ \citep{sembach2003}, the expected RM from such loops would be much smaller than the observed $|$RM$|$ of a few rad m$^{-2}$. In this case, such loops could not produce the RMs seen towards the caps.

\subsection{Large scale magnetic field in the halo of the Milky Way}

\label{section:largescalefield}
In this section, we consider the possibility that the observed RM pattern originates on the Galactic scale. We first estimate the implied magnetic field strength and direction from the observed RMs assuming that the observed Faraday rotation occurs in diffuse ionized gas. In reality, it is mostly in the WIM that the Faraday rotation takes place because even with a higher filling factor, the hot halo electrons ($\sim$ 10$^5$ $-$ 10$^6$ K) are of very low densities ($\le$ 10$^{-3}$ - 10$^{-4}$ cm$^{-3}$). Therefore, it is reasonable to assume that the hot halo electrons have negligible contributions to the observed RMs. We then consider the possibilities that the observed magnetic field is the result of either a Galactic wind, a large scale Galactic dynamo or a primordial field.

\subsubsection{Vertical Magnetic Field}
\label{subsection:vertical}
After the removal of extreme RMs and anomalous RM regions from the data set following the method described in \S~\ref{section:results}, one can infer the properties of the Galactic magnetic field from the remaining EGSs RMs assuming that the WIM is where most of the Faraday rotation takes place.

The observed RM is the integral of the projection of the Galactic magnetic field along the line of sight weighted by thermal electron density. At the highest latitude, RMs measure mostly the vertical component of the Galactic magnetic field, as the projection of the horizontal component along the line of sight is very small. In a right handed coordinate system centered at the location of the Sun where the positive $z$  axis points from the south Galactic pole to the north Galactic pole and the positive $x$ axis points towards the Galactic center, the Galactic magnetic field is of the form
\begin{equation}
\vec{B} = B_H \cos (l_0) \hat{x} +B_H \sin(l_0) \hat{y}+B_z \hat{z}, 
\end{equation}
 where $B_z$ is the coherent component perpendicular to the Galactic disk (defined such that a positive $B_z$ implies a field pointing from the south to the north Galactic pole), and $B_H$ is the coherent component parallel to the Galactic disk directed along Galactic longitude $l_0$. We consider $B_z$, $l_0$ and $B_H$ to be constant  within the 13$^\circ$ radius cone that we are probing around the Galactic poles. 
If we use the definition of RM (Eq~\ref{eq:rmdef}) and assume that the volume averaged thermal electron density $n_e$ in the Milky Way is an exponential disk of mid-plane density $n_{e,0}$ and scale height $H_0$ :
\begin{equation}
\label{eq:ne}
n_e(z) = n_{e,0} e^{-|z|/H_0} ,
\end{equation}
then the RM towards Galactic coordinates ($\ell$,$b$) can be expressed as
\begin{equation}
\label{eq:rmmodel}
{\rm RM} = 0.812 n_{e,0} H_{0} (a B_z - B_H \cos(\ell-\ell_0)/\tan|b|),
\end{equation}
where $a$ =+1 for $b$$<$$-$77$^\circ$ and $a$= $-1$ for $b$$>$+77$^\circ$.

If we take the average RM, $ \langle {\rm RM} \rangle$, along many sight lines with full coverage in $\ell$ towards the caps,  the contribution from the horizontal field vanishes. The vertical magnetic field strength is related to the integrated thermal electron column density towards the Galactic poles, ${\rm DM_\perp}$=$n_{e,0} H_0$ by 
\begin{equation}
\label{eq:bz}
B_z = \frac{ \langle {\rm RM} \rangle } {0.812 a \rm {DM_\perp}}. 
\end{equation}
We note that while there remain controversies on the exact values of the scale height $H_0$ and the mid-plane electron density $n_{e,0}$, the total column density of thermal electrons is well constrained to be $\sim$ 25 pc cm$^{-3}$ using pulsar DMs at high $z$ \cite[see for example,][]{cl2003,gaensler2008}. 

From Eq~\ref{eq:bz}, we found a vertical magnetic field towards the north Galactic pole to be consistent with zero (0.00  $\mu$G $\pm$  0.02 $\mu$G), with a 3$\sigma$ upper limit on the vertical magnetic field strength of  0.07 $\mu$G. On the other hand, the vertical magnetic field towards the south Galactic pole is found to be  $+$0.31 $\mu$G $\pm$ 0.03 $\mu$G. We note that the reduced $\chi^2$s of a model with the derived vertical magnetic field strengths towards the Galactic poles exceed unity ($\chi_r^2$ $\sim$ 7.9 towards the north and $\chi_r^2$ $\sim$ 10.6 towards the south). This is because systematic RM scatter is introduced by intrinsic RMs of EGSs and small-scale Milky Way foreground electron density and magnetic field fluctuations. 

\cite{wu2009} have found, using isothermal magnetohydrodynamic turbulence simulation, a relation between the distribution of normalized RM and the line of sight magnetic field strength. One can estimate the vertical magnetic field strength towards the poles using Equation (3) of \cite{wu2009} and the median and standard deviation of the RM data set towards the north and the south caps reported in \S~\ref{subsection:rmdist}. The estimated vertical magnetic field strength towards the north and south Galactic pole are roughly 0 $\mu$G and  0.46 $\mu$G respectively, which agree in general with our results. However, as \cite{wu2009} have pointed out, the relation is valid only for  a Mach number of unity and thus it is unclear if it holds in the diffuse ionized gas at high Galactic latitude.

There are two pulsars with measured RMs towards the Galactic caps: PSRs J0134-2937  and B1237+25 \citep{taylor1993,han1999}. Their coordinates, DMs and RMs are listed in Table~\ref{table:pulsar}. One can estimate the vertical magnetic field between the Sun and the pulsar using DM$_{\rm pulsar}$ and RM$_{\rm pulsar}$
\begin{equation}
\label{eq:pulsarbz}
B_z = \frac { {\rm RM_{pulsar}} \sin{|b|}} {0.812 a {\rm DM_{pulsar}}}, 
\end{equation}
where $b$ is the Galactic latitude of the pulsar. The vertical magnetic field derived from B1237+25 is $+$0.044 $\pm$ 0.008 $\mu$G, while that derived from J0134-2937 is +0.7 $\pm$ 0.1 $\mu$G. The vertical magnetic field strength derived using the pulsars roughly agree with results obtained using RMs of EGSs. Since PSR J0134-2937 has a much higher DM than PSR B1237+25, it probes through a longer path length in the WIM than the northern pulsar.

\cite{taylor2009} reported a vertical field of $-$0.14 $\mu$G $\pm$ 0.02 $\mu$G towards the north Galactic pole and +0.30 $\mu$G $\pm$ 0.03 $\mu$G towards the south Galactic pole. 
While our measured vertical field towards the south Galactic cap is consistent with \cite{taylor2009} within errors, our estimation of the vertical field towards the north Galactic cap disagrees with \cite{taylor2009}. This is likely due to the fact that \cite{taylor2009} have averaged over a larger region around the north Galactic pole without discarding outliers and anomalous RM regions around the pole before performing the fit. Different individual RMs derived from the NVSS catalogue and our WSRT observations due to multi-RM component sources and different ionospheric conditions (\S~\ref{subsection:rmcompare}) might also contribute to this discrepancy.

\subsubsection{Horizontal Field}
\label{subsection:horizontal}

In \S~\ref{subsection:vertical}, we used the fact that the median RM of EGSs is zero to reach the conclusion that there is no vertical field towards the north Galactic pole. However, it is still possible that there exists a horizontal magnetic field at high positive Galactic latitude. We can test this by fitting the north Galactic cap RMs to a model with only a horizontal component\footnotemark[12]\footnotetext[12]{This is obtained by setting  $B_z$ = 0 $\mu$G in Eq~\ref{eq:rmmodel}.}
\begin{equation}
\label{eq:rmmodelhorizontal}
{\rm RM} = -0.812 {\rm DM_\perp} (B_H \cos(\ell-\ell_0)/\tan|b|), 
\end{equation}
minimizing the $\chi^2$ between the observed RMs and the modeled RMs predicted by the above equation.
The best fit parameters are $B_H$=$0.6_{-0.4}^{+0.8}$ $\mu$G and $\ell_0$= $153^{\circ{+53^\circ}}_{~-101^\circ} $ with a reduced $\chi^2$ of 7.7. Comparing this model to a model with no magnetic fields using the F-test, the significance of a horizontal field of strength 0.6 $\mu$G is only at  2.4 $\sigma$ level.  This is not surprising as no obvious sinusoidal variation of RM as a function of $\ell$ could be seen in the smoothed RM map in the top panel of Fig~\ref{fig:smooth_rm}. Both the derived horizontal field strength and its direction differ from the best fit values towards positive mid-latitude (0.39 $\mu$G at $\ell_0$= 281$^\circ$) obtained by \cite{taylor2009}. As illustrated in Figure 7 of \cite{taylor2009} , the best fit parameters obtained when fitting to high positive latitude RMs are different. The authors attributed this to a potentially more complicated halo magnetic field. 

We can also test if there is a horizontal magnetic field component in addition to the detected vertical component  towards the south Galactic pole by performing a least square fit to a Galactic magnetic field model with both a vertical and a horizontal component (Eq~\ref{eq:rmmodel}). The reduced $\chi^2$ of the best fit to such a model is 10.6. The F-test suggests the existence of an additional horizontal field to the measured vertical field is significant at only 1.7 $\sigma$ level. This is expected because no sinusoidal RM variations in $\ell$ could be seen in the bottom panel of Fig~\ref{fig:smooth_rm}. We conclude that there is little evidence for a horizontal field towards the Galactic poles.

Instead of attempting to fit for the horizontal field, we can subtract its contribution from the measured RMs using the best fit values obtained by \cite{taylor2009} to check if it changes our estimation of the vertical magnetic fields in \S~\ref{subsection:vertical}. This is justified only if the halo magnetic field at high latitude is the same as that at mid-latitude. At $|$$b$$|$=77$^\circ$, the horizontal component contributes a maximum $|$RM$|$ of $\sim$ 1.9 rad m$^{-2}$  towards the north Galactic pole and 4.3 rad m$^{-2}$ towards the south Galactic pole. We have subtracted the contribution of the \cite{taylor2009} horizontal halo magnetic field from our measured RMs and found that it does not alter the vertical magnetic field estimates presented in \S~\ref{subsection:vertical} within the errors.

 \subsection{Galactic Wind}
X-shaped polarization pattern observed in halos of nearby edge-on galaxies imply large vertical magnetic fields increasing with height above the galactic disk \citep{beck2008b}. A kinematic disk dynamo, which generates dominant toroidal magnetic fields, cannot alone explain the existence of these fields. As some of these galaxies exhibit evidence of cosmic ray driven  winds, an alternative explanation of the large vertical field is wind advection that transports magnetic fields from the disk into the halo, distorting the expected field structure from dynamo actions \citep[see for example, ][]{heesen2009}. 

Recent studies by \cite{everett2008,everett2009} have successfully reproduced the Milky Way's diffuse soft X-ray emission and synchrotron emission using a 1D thermally and cosmic ray driven wind model for the Galaxy. In their cosmic ray driven wind model, the wind is launched within a Galactocentric radius of $\sim$ 4.5 kpc, but flares to larger Galactocentric radii above and below the plane, and hence there is no wind launched at the location of the Sun. Similarly, \cite{breitschwerdt1991,breitschwerdt1993} have argued that a wind launched from the solar neighborhood would lead to too much cosmic ray escape and hence inconsistent with the inferred residence time of cosmic rays in the Galaxy. \cite{everett2008,everett2009} have adopted a flared-cylinder wind geometry where the wind stays well confined within a cylinder of constant cross-section up to a height z$_{break}$ above/below the Galactic plane. Beyond z$_{break}$, the cross-sectional area of the wind increases as a power law of z. At some z, this wind is directly above/below the location of the Sun. However, the best fit value of z$_{break}$ is found to be $\sim$ 4 kpc \citep{everett2009}. At this height above/below the Galactic plane, the density of the exponential WIM  is too low to produce any observable Faraday rotation. Therefore, we conclude that the observed RM towards the Galactic caps is unlikely to be due to a large scale Galactic wind.

\subsection{Mean Field Dynamo Theory} 
\label{subsection:dynamo}

The existence of galactic-scale coherent magnetic fields in the Milky Way disk and in other normal spiral galaxies with significant differential rotation can be explained by the standard mean field $\alpha$-$\omega$ dynamo \citep{beck1996}, although the theory and its application to galaxies has been questioned on theoretical grounds \citep[see for example][for a recent summary]{cattaneo2009}. On a time scale of a few Gyrs, this process amplifies and orders the field by turbulence rising into the halo, transforming an azimuthal field into a poloidal one (the $\alpha$-effect) and by differential rotation in the underlying disk, transforming the radial component of the poloidal field back into an azimuthal one (the $\omega$-effect) \citep{shukurov2004}.  In this section, we examine if dynamo theory can predict the strength and geometry of a vertical field that is consistent with the values reported in  \S~\ref{subsection:vertical}.

The  $\alpha$-$\omega$ dynamo predicts an azimuthal field that dominates over the vertical field by more than a factor of 10 because the $\omega$-effect operates more efficiently than the $\alpha$-effect \citep{ferriere2005}. We can crudely estimate the expected local vertical magnetic field strength by evoking $\nabla$ $\cdot$ $\vec{B}$ = 0 using a local horizontal field of $\sim$ 2 $\mu$G at a pitch angle of 15$^\circ$ \citep{beck1996}, and a Galactic disk of total thickness 2$h$  and diameter 2$R$ (where  $h$ $\sim$ 2 kpc and $R$ $\sim$ 15 kpc). The estimated vertical magnetic field strength at the location of the Sun is $\sim$ B$_{radial}$ ($h$/$R$) $\sim$ 0.07 $\mu$G. Another estimation of this ratio follows from dynamo theory:  B$_z$ $\sim$ B$_{radial}$ $\sqrt{h/R}$ \citep{ruzmaikin1988}. The expected vertical magnetic field strength is $\sim$ 0.2 $\mu$G, which is similar to the vertical magnetic field strength we obtained for the south Galactic cap. The dynamo mechanism is thus capable of producing the observed vertical magnetic field strength. 

The large scale magnetic field configuration of a galaxy can be classified by axial symmetry with respect to the rotation axis as well as by vertical symmetry with respect to the galactic mid-plane. For a strongly differentially rotating galaxy, a classical mean-field dynamo favors even field symmetry (a quadrupolar field), in which the toroidal component is symmetric across the mid-plane while the vertical component reverses direction \citep{shukurov2004}. For a weakly differentially rotating galaxy, or a halo, odd field symmetry (a dipolar field) is preferred, in which the toroidal component reverses direction across mid-plane while the vertical component does not \citep{ferriere2005}.  However, numerical simulations carried out  by \cite{ferriere2000} found comparable growth rates of the odd and the even symmetry modes, suggesting that the present Galactic magnetic field might be of mixed parity rather than of pure even or odd parity. As mentioned in \S~\ref{section:introduction}, studies of the Milky Way's field parity have yielded diverging results, possibly because they focus on different Galactic latitude ranges and because it is difficult to distinguish local magneto-ionic effects from a genuine large scale field. The most convincing piece of work is the wavelet analysis of all sky RMs conducted by \cite{frick2001}. They found that the local magnetic field has an even parity -- the horizontal field component does not reverse direction across the Galactic plane. If the Milky Way's large scale field is indeed quadrupolar in nature, then one expects the vertical magnetic field to reverse direction across the mid-plane. On the other hand, \cite{han1994} and \cite{han1997,han1999} concluded from an all sky smoothed RM map that the RM distribution  towards the inner Galaxy is anti-symmetric across the Galactic plane. The authors attribute this to a large scale dipolar field. If this is the case, then one expects the vertical magnetic field to have the same direction above and below the Galactic plane. In \S~\ref{subsection:vertical}, we found that the vertical field is consistent with zero for $b$$>$ +77$^\circ$ and that $B_z$ =$+$0.31 $\pm$ 0.03 $\mu$G for $b$$<$-77$^\circ$. This is inconsistent with either a quadrupole or a dipole large scale field configuration. Even though a simple disk dynamo can produce a vertical field strength comparable to that being observed towards the south cap, it cannot explain the observed vertical field geometry. 

\cite{sokoloff1990} demonstrated that the mean field dynamo can operate in the halo of a rotating galaxy since the mean helicity (the $\alpha$-effect) of gaseous halos are non-zero due to galactic fountains/Parker instabilities. The dominant mode is an axisymmetric field with odd parity with respect to the Galactic disk; this is of the opposite parity from the mode excited by the dynamo action in the Galactic disk. The asymmetric RM pattern that we see  towards the north/south Galactic hemisphere seems to support this idea. In particular, the shift of RM structure on the largest scales to $b$=$-$15$^\circ$ found by \cite{frick2001} can be explained by a dipolar field in the halo of the Galaxy such that the vertical magnetic field from the disk and the halo add up in the southern hemisphere. \cite{sun2008} were able to obtain a reasonable fit to the latitude extension of CGPS measurements using an asymmetric halo field plus a symmetric disk field. More recently, \cite{taylor2009}  found that mid-latitude NVSS RMs can be well fitted with a $\sim$ 0.4 $\mu$G toroidal halo field that reverses direction across the mid-plane. If indeed an anti-symmetric halo field and a symmetric disk field co-exist in the Milky Way, then on the side of the Galactic disk (in this case $b$$>$0$^\circ$), the vertical component of the halo field and that of the disk field would partly cancel out, as they are oppositely directed. These two vertical components would add up on the other side of the disk ($b$$<$0$^\circ$). Depending on the relative vertical field strength between the halo/disk field and their extension above/below the disk, it is not impossible that they can cancel each other out exactly towards the northern hemisphere resulting in a net RM that is consistent with zero. 

However, this theory has been challenged by numerical work. \cite{brandenburg1992} showed that the halo dynamo requires more than a Hubble time to reach the steady state configuration and thus the field that one observes at this moment might merely be a transient field of mixed parity. \cite{moss2008} solved the mean field dynamo equations for a system with a disk and a halo dynamo. These authors were not able to produce any co-existing system of a  dipole-like halo field and a quadrupole-like disk field; instead, one always dominates over the other, though \cite{moss2008} acknowledged that using a larger turbulence diffusivity ratio between the halo and the disk might mitigate the problem. 

We conclude that the observed vertical magnetic field geometry towards the Galactic poles is not consistent with predictions from a pure disk dynamo. A separate halo dynamo of odd parity could potentially account for the observed vertical magnetic fields at high latitude but until now no numerical simulation has successfully produced a co-existing disk and halo fields of opposite parity in a steady state.

\subsection{Primordial origin}
The competing theory to the dynamo origin of galactic magnetic fields is that of primordial origin. The primordial field theory suggests the following: if the IGM field is frozen into the gas, then the total magnetic field would be enhanced by a few orders of magnitude as gas clouds collapse to form a protogalaxy. This relic field would then be modified by the differential rotation of the galaxy, producing the present day galactic magnetic fields  \citep[see for example,][]{beck1996,howard1997}. The component of the seed field parallel to the galactic disk can be removed diffusively and through large scale flow, but the component of the magnetic field parallel to the rotation axis of the galaxy is trapped \citep{ruzmaikin1988}.

 Since galactic rotation is symmetric with respect to the plane, one expects the azimuthal component of such a field to reverse its direction across the mid-plane while the vertical component preserves its direction, thus resulting in a dipolar type field \citep{ruzmaikin1988,beck1996}. The strength of a vertical galactic magnetic field of primordial origin depends highly on the initial orientation of the intergalactic seed field with respect to the rotation axis. The collapsed gas that forms the proto-galaxy could potentially increase its strength by two orders of magnitude \citep{kulsrud2008}. This small field is preserved until the present day since the total vertical magnetic flux is conserved. The concept of a primordial field may be overly simplistic because galactic disks probably build up over time through mergers and infall. The addition of new material can add new magnetic flux, but whatever flux is added is subject to the constraints described above. 

The observed vertical field geometry is not consistent with that from a pure dipole field of primordial origin: the vertical field from the south Galactic pole is directed toward us, while the vertical field toward the north Galactic pole is consistent with zero. Therefore, we cannot attribute the observed vertical magnetic field to a primordial field alone.

\section{Turbulence and Random Magnetic Field at high Galactic latitude}
\label{section:random}
 In \S~\ref{subsection:rmsf}, we have shown that the RM SFs  are flat, implying that no specific scale is associated with the observed RM pattern towards the north and the south Galactic caps. In this section, we will place limits on turbulence properties at high $|b|$ as well as estimate the random magnetic field strength using a plane-parallel free electron density model.
 
If we assume that the observed RM of an EGS is the sum of only its intrinsic RM and the RM produced by the magneto-ionic medium in the Milky Way, then the observed structure function is the superposition of the intrinsic RM structure function and that of the foreground. The SF of intrinsic RMs of EGSs is expected to be flat, because intrinsic RM of independent EGSs are uncorrelated with each other, at a saturation level of  2$\sigma_{\rm RM,intrinsic}^2$. On the other hand, a turbulent Galactic foreground with an outer scale  $l_{outer}$ and  and an inner scale $l_{inner}$ would produce a SF that rises between $l_{\rm inner}$  and $l_{\rm outer}$ and saturates beyond $l_{outer}$ at  2$\sigma_{\rm RM,foreground}^2$, where $\sigma_{\rm RM,foreground}$ is the variance of RM due to random magnetic fields or variations in electron density in the foreground ISM. Fig~\ref{fig:rmsf} shows that the observed RM structure function towards both Galactic caps saturates at roughly 170 rad$^2$ m$^{-4}$, which corresponds to $\sigma_{\rm RM,total}$ $\sim$ 9 rad m$^{-2}$. This is then also the upper limit of $\sigma_{\rm RM,foreground}$.

The flatness of the observe RM SFs suggests that intrinsic RM of EGSs dominate the SF, although one cannot rule out the possibility that the outer scale of turbulence in the Galactic ISM is below the smallest angular scale probed. This result is consistent with previous turbulence studies towards the north Galactic pole carried out by \cite{simonetti1984} and \cite{sun2004} who both found flat RM structure functions on scales greater than 1$^\circ$-2$^\circ$ and concluded that the intrinsic EGS RM dominates the SF. However, they have reported a saturation level at log(SF$_{\rm RM}$) $\sim$ 2.8 which implies a larger total RM standard deviation ($\sim$ 16 rad m$^{-2}$) than what we have measured. This discrepancy is likely due to the different methods used to handle the RM measurement errors in computing the SF. It is unclear if the SF of RM measurement errors was subtracted from the measured SF in these previous works. Also, RM outliers in the sample and inaccurate RMs can lead to offsets in the SF. We note that the SFs computed by \cite{simonetti1984} and by \cite{sun2004} were all based on a relatively small number of EGSs (52  sources at $b$ $>$ +60$^\circ$ in \cite{simonetti1984} and 35 sources at $b$ $>$ +70$^\circ$ in \cite{sun2004}), sparsely sampled towards a large region near the pole.

Next, we use the coherent vertical magnetic field strength derived in \S~\ref{subsection:vertical} and properties of turbulence in the Galactic halo to estimate the random magnetic field strength at high Galactic latitude. We construct a  cell model similar to those used to estimate the random field in the inner Galaxy \citep{gaensler2001}, the Large Magellanic Cloud \citep{gaensler2005} and the Small Magellanic Cloud \citep{mao2008}.

We assume that the WIM is composed of plane-parallel cells of size $l_{\rm outer}$ in pc, equivalent to the outer scale of turbulence. We also assume that all the sight lines towards EGSs in a 13$^\circ$ cone around the Galactic poles are parallel. We introduce a clumpy WIM, where the volume average electron density $n_e(z)$ (Eq~\ref{eq:ne}) is the product of the filling factor (the fraction of the line of sight at height $z$ filled with electrons) and the internal cell electron density $N(z)$. We have adopted the evolution of the filling factor and the cell electron density with height above the Galactic plane by \cite{gaensler2008}. 

\begin{equation}
\label{equation:ffcd}
f(z)=f_0 e^{+|z|/H_f}, 
\end{equation}
\begin{equation}
\label{equation:nc}
N(z)=N_0 e^{-|z|/H_N},
\end{equation}
where $f_0$=0.04  and $N_0$=0.34 are the mid-plane filling factor and internal cell density;  $H_f$ $\sim$ 700 pc and $H_N$ $\sim$ 510 pc are the scale heights of the filling factor and the internal cell density respectively. Eq~\ref{equation:ffcd} and ~\ref{equation:nc}  were derived by \cite{gaensler2008} using independent constrains on $n_e(z)$ from pulsar DMs and on  $n_e^2(z)$ from WHAM H$\alpha$ measurements towards the Perseus arm. For  $z>$ 1.4 kpc, beyond the height of WIM observed by WHAM towards the Perseus arm, the above equations are no longer valid as  $n_e^2(z)$ is not constrained. Below $z$ of 1.4 kpc, the filling factor increases exponentially with height as one expects the WIM to dominate over the neutral component at higher $z$. In each cell, we assume that there is a constant coherent magnetic field $B_z$$\sin$$b$ $\approx$ $B_z$ (for $b$ $\approx$ 90$^\circ$) parallel to the line of sight and a magnetic field with constant strength $B_{ran}$ but random orientations $\theta$ with respect to the line of sight in different cells. For simplicity, we assume that $B_z$ and $B_{ran}$ are independent of $z$. The line of sight magnetic field through a particular cell $m$ is given by 
\begin{equation}
B_{cell,m} = B_z+B_{ran} \cos \theta. 
\end{equation}
The average RM across many random sight lines through $M$ cells is given by
\begin{equation}
\label{eq:avgrm}
\langle {\rm RM}_{ M cells} \rangle = 0.812 B_z \int_{M l_{outer}}^0 n_e(z) dz, 
\end{equation}
where $M*l_{\rm outer}$=1.4 kpc. On the other hand, the average of squared RM over many random sight lines depends explicitly on the cell electron density and the filling factor.
\begin{align}
\langle {\rm RM}_{M cells}^2  \rangle = &0.812^2 (B_z^2 +\frac{1}{3} B_{ran}^2) l_{outer} \int_{M l_{outer}}^0 f(z)N^2(z) dz~+ \notag \\ 
 & 0.812^2 B_z^2 \{ \left[\int_{M l_{outer}}^0 n_e(z) dz\right]^2 -l_{outer} \int_{M l_{outer}}^0n_e^2(z) dz\}   \label{eq:avgrmsq} 
\end{align}

Using Equation~\ref{eq:avgrm}, \ref{eq:avgrmsq} and the definition of standard deviation, one can write down the expression of the random magnetic field strength in the WIM in terms of the standard deviation of RM arised in the foreground magneto-ionic medium, $\sigma_{\rm RM,foreground}$
\begin{equation}
\label{eq:randomfield2}
B_{ran} = \sqrt{ 3 \left[\frac{\sigma_{\rm RM,foreground}^2 + 0.812^2 B_z^2 l_{outer} \int_{M l_{outer}}^0 n_e^2(z) dz  } {0.812^2 l_{outer} \int_{M l_{outer}}^0 f(z)N^2(z) dz }-B_z^2\right]}~.
\end{equation}

We have placed an upper limit to the  the standard deviation of RM due to the Galactic foreground $\sigma_{\rm RM,foreground}$ of  approximately 9 rad m$^{-2}$ earlier in this section.  We assume an outer scale of turbulence   ($l_{\rm outer}$) of $\sim$ 90 pc estimated using the autocorrelation function of synchrotron radiation measured towards the north Galactic pole \citep{lazaryan1990}. As Eq~\ref{equation:ffcd} and \ref{equation:nc} are physically meaningful only for $z$$<$1.4 kpc, we choose to carry out the integral in Eq~\ref{eq:randomfield2} to $z$ =  $M*l_{\rm outer}$ = 1.4 kpc. Using a uniform magnetic field strength $B_z$ = 0 $\mu$G, we find that $B_{ran}$ $\le$1.5 $\mu$G towards the north Galactic pole. Assuming that turbulence towards the south Galactic pole has the same outer scale as that towards the north and using a uniform vertical field of $|B_z|$ = 0.31 $\mu$G, we obtain an upper limit of the random magnetic field strength of 1.4 $\mu$G towards the south Galactic pole. Our estimated random field strength of $\sim$ 1 $\mu$G is systematically smaller than other estimates of the random field in the Milky Way disk: 4-6 $\mu$G was inferred from  DM and RM of pulsars \citep{ohno1993} and  5 $\pm$ 2 $\mu$G inferred using the Haslam 408 MHz synchrotron intensity map \citep{beck1996,shukurov2004}. If one assumes energy equipartition between the turbulence and random magnetic fields, $B_{ran}$ is approximately 
\begin{equation}
B_{ran} = \sqrt{4 \pi \rho v^2},  
\end{equation}
where $v$ is the rms turbulent velocity and $\rho$ is the density \citep{beck1996}. Since both the density and the turbulent velocity in the halo is smaller than that in the Galactic disk, the random magnetic field in the halo is expected to be smaller than the values estimated for the Galactic plane. \cite{cox2005} assumed equipartition between thermal and non-thermal pressure and derived a field strenth of $\sim$ 2 $\mu$G at a few kpc above/ below the Galactic plane. This is consistent with the $\sim$ 1$\mu$G random magnetic field that we computed.

\section{Summary and Outlook}
\label{section:conclusions}

In this paper, we have presented an RM survey with the ATCA and the WSRT of polarized extragalactic radio sources towards the north and the south Galactic poles at $|b|$ $\ge$ 77$^\circ$. Using rotation measure synthesis, we have obtained 813 reliable RMs towards the Galactic poles.  No preferred RM fluctuation scale was apparent from the flat RM structure functions towards the Galactic caps. After discarding outliers and anomalous RM regions in \S~\ref{section:results}, we obtain a median RM of 0.0 $\pm$ 0.5 rad m$^{-2}$ towards the north Galactic pole; and a median RM of +6.3 $\pm$ 0.7 rad m$^{-2}$ towards the south Galactic pole.

In \S~\ref{subsection:local}, we have ruled out the possibility that local sources/events such as the LISM, the LB and a Parker's instability loop produce the observed RM pattern. In \S~\ref{subsection:vertical} and~\S~\ref{subsection:horizontal}, we have derived the halo magnetic field properties from the observed RMs assuming that they are produced by the diffuse interstellar free electrons. We found no evidence for  vertical and horizontal magnetic field towards the north Galactic pole. On the other hand, a vertical field of strength +0.31 $\pm$ 0.03 $\mu$G  was detected at $>$9 $\sigma$ towards the south Galactic pole, but there is no evidence for an additional horizontal component. Although a dynamo or a primordial field can explain the derived vertical magnetic field strength towards the Galactic poles, a pure dipole/ quadrupole field cannot explain the geometry of the observed vertical field across the mid-plane. One possible explanation of the derived magnetic field properties is that proposed by \cite{sokoloff1990}, in which a disk and a halo dynamo of different parities are simultaneously at work in the Galaxy. This could potentially lead to part cancellation of RM produced by the vertical magnetic field in the northern Galactic hemisphere, which is compatible with our RM measurements. However, until now, no numerical simulation has been able to produce a co-existing system of a  dipole-like halo field and a quadrupole-like disk field. Numerical works that explore larger parameter space (especially turbulent diffusivity) is needed to test this hypothesis.  Finally, we have estimated the random magnetic field strength in the halo of the Milky Way by constructing a plane-parallel cell model of the WIM and the standard deviation of RMs  to derive a random magnetic field strength of $\sim$ 1 $\mu$G in the Galactic halo, which is smaller than the random field in the mid-plane of the Galaxy, but in equipartition with the lower turbulent energy density inferred for the halo.

Exploration of cosmic magnetism is one of the key sciences of the next generation radio telescopes -- the Square Kilometre Array (SKA) and its prototypes such as the Australian Square Kilometre Array Pathfinder (ASKAP), which are capable of providing accurate RMs of EGSs densely sampled over the entire sky \citep{johnston2007,johnston2008}. For example, one of the approved Survey Science project of ASKAP -- the Polarization Sky Survey of the Universe's Magnetism (POSSUM) aims to perform RM synthesis and obtain a grid of RMs over a large fraction of the sky. Similar projects will provide a more detailed picture on the magnetic field  and turbulence properties at high Galactic latitude. Also, pulsar searches at high Galactic latitudes will allow one to probe the vertical magnetic field as a function of height above/ below the Galactic disk by simultaneously using pulsar DM and RM, which can further constrain the structure of the Galactic halo magnetic field.

\textbf{Acknowledgements}

We thank Eve Meyer and Gemma Anderson for helping to carry out the ATCA observations; Observer's friend Ger de Bruyn and Gyula Jozsa to help with the preparation of the WSRT observations. Robert Braun for helping with the WSRT data calibration with AIPS, Douglas Finkbeiner for useful discussions, and Justin Kasper for detailed discussion on the ionospheric rotation measure correction. 
This work was supported in part by an Australian Research Council Federation Fellowship (FF0561298) awarded to B. M. G.. The Wisconsin H-Alpha Mapper is funded by the National Science Foundation. The Southern H-Alpha Sky Survey Atlas (SHASSA) is supported by the National Science Foundation. The Westerbork Synthesis Radio Telescope is operated by the ASTRON (Netherlands Institute for Radio Astronomy) with support from the Netherlands Foundation for Scientific Research (NWO). The Australia Telescope Compact Array is part of the Australian Telescope, which is funded by the Commonwealth of Australia for operation as a National Facility managed by CSIRO.
 
 {\it Facilities:} ATCA WSRT

\clearpage

\clearpage

\clearpage

\clearpage

\clearpage

\clearpage

\clearpage
\clearpage

\clearpage
\clearpage

\clearpage
\clearpage

\clearpage
\LongTables
\begin{landscape}
\begin{deluxetable}{lllrrrrc}
\tablecolumns{8} 
\tabletypesize{\footnotesize}
\tablewidth{0pc} 
\tablecaption{Rotation Measures of Extragalactic Sources towards the North Galactic Pole, sorted by Galactic longitude} 
\vspace{0.01in}
\tablehead{   
 \colhead{RA(J2000)(hms)} &     \colhead{DEC(J2000)(dms)}   & \colhead{$\ell$} & \colhead{$b$} &     \colhead{RM(rad m$^{-2}$)}   & \colhead{PI (mJy)\tablenotemark{*}} & \colhead{I (mJy)\tablenotemark{*}} & \colhead{Used in \S~\ref{section:largescalefield}?} }
%\label{table:northrms}
\startdata 
13:07:08.33 & +24:47:00.70 &  0.21 & +85.76 & $-$3 $\pm$ 4 & 7.77 & 131.49 & Yes \\
13:35:48.14 & +20:10:16.00 &  0.86 & +77.70 & $+$3 $\pm$ 5 & 8.72 & 71.47 & No\tablenotemark{b} \\
13:24:14.48 & +22:13:13.10 &  1.33 & +81.08 & $-$6 $\pm$ 6 & 7.62 & 148.72 & Yes \\
13:24:19.18 & +22:14:00.50 &  1.47 & +81.07 & 0 $\pm$ 6 & 8.16 & 104.47 & Yes \\
13:16:04.36 & +23:37:37.70 &  2.10 & +83.43 & $+$4 $\pm$ 13 & 3.88 & 25.64 & Yes \\
13:20:48.54 & +22:57:54.70 &  2.49 & +82.16 & $-$6 $\pm$ 8 & 4.81 & 60.32 & Yes \\
13:11:16.33 & +24:27:15.60 &  3.11 & +84.80 & $+$5 $\pm$ 3 & 10.14 & 90.23 & Yes \\
13:34:25.53 & +21:10:04.50 &  3.93 & +78.53 & $-$1 $\pm$ 9 & 4.84 & 183.10 & No\tablenotemark{b} \\
\enddata
\label{table:northrms}
\tablenotetext{a}{The source is discarded as it lies within 2$^\circ$ from the Coma Cluster.}
\tablenotetext{b}{The source is discarded as its position coincides with the North Polar Spur region.}
\tablenotetext{c}{The source is discarded as it is 3.2$\sigma$ away from the median of the RM distribution.}
\tablenotetext{d}{The source is discarded as it lies within an anomalous RM region, indicated in Fig~\ref{fig:smooth_rm}.}
\tablenotetext{*}{The debiased polarized flux and total intensity are estimated at the location of the brightest polarized pixel at 20 cm.}

\end{deluxetable} 
\clearpage
\end{landscape}
\clearpage

\clearpage
\LongTables
\begin{landscape}
\begin{deluxetable}{lllrrrrc}
\tablecolumns{8} 
\tabletypesize{\footnotesize}
\tablewidth{0pc} 
\tablecaption{Rotation Measures of Extragalactic Sources towards the South Galactic Pole, sorted by Galactic longitude} 
\vspace{0.01in}
\tablehead{   
 \colhead{RA(J2000)(hms)} &     \colhead{DEC(J2000)(dms)}   & \colhead{$\ell$} & \colhead{$b$} &     \colhead{RM(rad m$^{-2}$)}   & \colhead{PI (mJy)\tablenotemark{*}} & \colhead{I (mJy)\tablenotemark{*}} & \colhead{Used in \S~\ref{section:largescalefield}?}    }
%\label{table:southrms}
\startdata 00:11:13.31 & -31:50:31.20 &  2.21 & -80.07 & $+$8 $\pm$ 3 & 7.82 & 375.42 & Yes \\
00:09:25.06 & -31:46:03.60 &  3.52 & -79.75 & $+$23 $\pm$ 2 & 13.31 & 157.35 & Yes \\
23:58:31.56 & -32:38:23.40 &  4.06 & -77.29 & $-$33 $\pm$ 6 & 4.05 & 70.32 & No\tablenotemark{a} \\
00:09:45.21 & -31:27:01.80 &  5.01 & -79.94 & $-$1 $\pm$ 2 & 14.72 & 137.85 & Yes \\
00:14:37.96 & -30:59:18.80 &  5.11 & -81.08 & $+$9 $\pm$ 2 & 8.47 & 174.17 & Yes \\
00:10:34.68 & -30:54:19.80 &  7.59 & -80.31 & $+$7 $\pm$ 3 & 6.38 & 128.96 & Yes \\
23:56:40.83 & -31:49:23.20 &  8.15 & -77.20 & $+$4 $\pm$ 1 & 43.24 & 611.69 & Yes \\
23:59:35.47 & -31:33:45.90 &  8.40 & -77.87 & $+$5 $\pm$ 3 & 6.79 & 478.51 & Yes \\
\enddata
 \label{table:southrms}
 \tablenotetext{a}{The source is discarded as it is 3.2$\sigma$ away from the median of the RM distribution.}
\tablenotetext{*}{The debiased polarized flux and total intensity are estimated at the location of the brightest polarized pixel at 20 cm.}

\end{deluxetable} 
\clearpage
\end{landscape}
\clearpage

\begin{deluxetable}{lrrrrr}
\centering
\tablecolumns{6} 
\tablewidth{0pc} 
\tablecaption{Radio Pulsars with measured RMs at $|b|$ $>$ 77$^\circ$ } 
\tablehead{   
\colhead {Name} & \colhead{$\ell$} & \colhead{$b$}   & \colhead{DM (pc cm$^{-3}$)\tablenotemark{a}} &  \colhead{RM(rad m$^{-2}$)}
}

\startdata 
B1237+25 & 252.45 & 86.54 & 9.242 $\pm$ 0.006 & -0.33 $\pm$ 0.06 \tablenotemark{b}\\
J0134-2937  & 230.25 &  -80.25 & 21.806 $\pm$ 0.006 & 13 $\pm$ 2 \tablenotemark{c} \\

\enddata 
\label{table:pulsar}
\tablenotetext{a}{\cite{hobbs2004}}
\tablenotetext{b}{\cite{taylor1993}}
\tablenotetext{c}{\cite{han1999}}

\end{deluxetable}

\end{document}